\documentclass[preprint,showpacs,preprintnumbers,amsmath,amssymb]{revtex4}

\usepackage{graphicx}
\usepackage{dcolumn}
\usepackage{bm}

\begin{document}

\preprint{APS/123-QED}

\title{Neutrino oscillation and expected event rate of supernova
neutrinos in adiabatic explosion model}

\author{Shio Kawagoe}
\altaffiliation
{The University of Tokyo}
 \email{shiok@iis.u-tokyo.ac.jp}
\affiliation{
Knowledge Dissemination Unit, Institute of Industrial Science,
The University of Tokyo,\\
4-6-1 Komaba, Meguro-ku, Tokyo 153-8505, Japan\\
Division of Theoretical Astronomy, National Astronomical Observatory of Japan,
2-21-1 Osawa, Mitaka-shi, Tokyo 181-8588, Japan
}%
\author{Takashi Yoshida}
\affiliation{%
Department of Astronomy, School of Science,
the University of Tokyo,\\
 7-3-1 Hongo, Bunkyo-ku, Tokyo, 113-0033, Japan
}%
\author{Toshitaka Kajino}%
\affiliation{%
National Astronomical Observatory of Japan,
2-21-1 Osawa, Mitaka-shi, Tokyo 181-8588, Japan,\\
Department of Astronomy, School of Science,
the University of Tokyo,\\
 7-3-1 Hongo, Bunkyo-ku, Tokyo, 113-0033, Japan
}%
\author{Hideyuki Suzuki}
\affiliation{
Department of Physics, Faculty of Science and Technology,
Tokyo University of Science, \\2641 Yamazaki, Noda-shi,
Chiba 278-8510, Japan
}%
\author{Kohsuke Sumiyoshi}
\affiliation{
Numazu College of Technology, \\
3600 Ooka, Numazu-shi, Shizuoka 410-8501, Japan
}%
\author{Shoichi Yamada}
\affiliation{
Department of Physics, Faculty of Science and Engineering,
Waseda University, \\
3-4-1 Okubo, Shinjuku-ku, Tokyo 169-8555, Japan
}%

\date{\today}

\begin{abstract}

We study how the influence of the shock wave appears in
neutrino oscillations and the neutrino 
spectrum using density profile of adiabatic explosion model 
of a core-collapse supernova which is calculated in an
implicit Lagrangian code for general relativistic spherical 
hydrodynamics.
We calculate  expected event rates of neutrino detection at  
Super-Kamiokande and SNO 
for various $\theta_{13}$ values and both normal and inverted
 hierarchies.
The predicted event rates of $\bar{\nu}_e$ and  $\nu_e$ 
depend on the mixing angle $\theta_{13}$ for
the inverted and normal mass hierarchies, respectively,
 and 
the influence of the shock wave 
appears for about 2 - 8 s 
when $\sin^22\theta_{13}$ is larger than $10^{-3}$.
These neutrino signals for
the shock wave propagation is decreased by
 $\lesssim 30 \%$ for $\bar{\nu}_e$ in inverted hierarchy (SK) or by 
$\lesssim 15 \%$ for $\nu_e$ in normal hierarchy (SNO) compared with the
 case without shock.
The obtained ratio of the total event for high-energy neutrinos 
(20 MeV $\leqq E_{\nu} \leqq 60$ MeV) to low-energy neutrinos
(5 MeV $\lesssim E_{\nu} \leqq 20$ MeV)
is consistent with the previous studies
in schematic
semi-analytic or other hydrodynamic models of the shock propagation.
%
The time dependence of the calculated ratio of the
 event 
 rates of high-energy neutrinos 
to the
 event rates of low-energy neutrinos 
is a very
 useful observable which is sensitive to $\theta_{13}$
and mass hierarchies.
Namely,
 time-dependent ratio shows clearer signal of the shock wave propagation
 that exhibits remarkable decrease by at most factor $\sim 2$ for
 $\bar{\nu}_e$ in inverted hierarchy (SK), whereas
it exhibits smaller change by $\sim 10 \%$ for $\nu_e$ in normal
 hierarchy (SNO).
Observing time-dependent high-energy to low-energy ratio of the neutrino
 events thus 
would provide a piece of very useful information to constrain
$\theta _{13}$ and mass hierarchy, and eventually help
understanding the propagation
how the shock wave propagates inside the star. 

\end{abstract}

\pacs{Valid PACS appear here}
\maketitle

\section{\label{sec:level1}Introduction}

About twenty events of supernova neutrinos were detected from SN1987A 
\cite{hirata1987, bionta1987}. 
Some neutrino features such as neutrino temperatures and the energy
carried out by neutrinos 
that had been
obtained from the data of SN1987A were consistent with theoretical
expectation \cite{Sato1987, Suzuki1988, Suzuki1987, Suzuki1994}.
However, the detected number of the neutrino events was still
too small to find out more details 
about other flavors of
the supernova neutrinos and the explosion mechanism.
Although, we have not yet obtained 
neutrino events from the second core-collapse
supernova
which is the rare event of the century,
detection of next supernova neutrinos is highly desirable
to obtain important information on the 
neutrinos and the explosion mechanism.
The expected information 
from the supernova neutrino
observations is classified into two categories of implication for
neutrino physics and for supernova physics.

Large-volume underground detectors are now operating to detect various 
neutrino events.
For example,
Super-Kamiokande (SK), which is at the Kamioka mine in Japan, is a water
Cherenkov detector, filled with 50,000 ton pure water
(32,000
ton fiducial volume for the burst mode and 22,500 ton for the other
modes) 
\cite{Totsuka1992}.  
KamLAND is a liquid scintillator detector with 1,000 ton active volume
\cite{KamLAND1999}.  
Sudbury Neutrino Observatory (SNO) has operated as 
a  heavy water 
Cherenkov detector, filled with 1,000 ton fiducial
volume \cite{SNO2001}. 
Now, a new generation experiment, SNO+, is planned to be
 constructed \cite{kraus2006}.
We achieved remarkable development of the neutrino physics, especially
about neutrino oscillation from the solar, atmospheric and reactor
neutrino experiments with these detectors.
In addition, the supernova neutrinos are thought to be fascinating
targets 
of these detectors.
If one supernova explodes at the Galactic center, thousands of neutrino
events will be observed in Super-Kamiokande \cite{totani-sato-del}.
Furthermore, megaton-size neutrino detectors will detect 
$\times 10^5$ supernova neutrino events in the future 
\cite{ando2005, mosca2005, nakamura2003, titand, jung2000}. 
The neutrinos are released directly from the core, in which the
extreme  
 physical condition of very high density is realized. 
Therefore, the supernova neutrinos would become the probe of such an
environment. 

Neutrino flavor change by the neutrino oscillations relates to neutrino
oscillation parameters, i.e., the mixing angles, mass hierarchy, and CP
violation phase.
Most of neutrino oscillation parameters have been determined by the
various neutrino experiments
\cite{Fukuda2001, sno, Parameter}.
However, the mixing angle $\theta_{13}$ is not precisely constrained, and 
only the upper 
bound is known from reactor experiments (e.g., \cite{Apollonio}).
In addition, the neutrino mass hierarchy, 
i.e., the sign of $\Delta m^2_{13}\equiv m^2_3 - m^2_1$,
 and the CP violation phase remain
unknown. 
However, 
among many studies about implication on 
these unknown neutrino parameters from
future supernova neutrinos 
\cite{dighe, Fogli4, Lunardini03, Dighe03a, Takahashi3, Fogli2, Fogli3,
TAUP2007, nakazato2008},
there are several proposed possibilities that
the detection of the neutrinos from the next Galactic supernova would 
constrain the neutrino oscillation parameters and identify the mass
hierarchy more precisely \cite{Fogli2, Fogli3, TAUP2007, nakazato2008}. 

Most of the supernova neutrinos are released for about 10 seconds
after the core bounce,    
and interact with electrons when the neutrinos
propagate through stellar matter. 
Therefore, Mikheyev-Smirnov-Wolfstein (MSW) effect
is to be taken into account in the neutrino
oscillations of the supernova neutrinos. 
The MSW  effect on the supernova neutrino 
signal has been investigated previously in literatures (e.g., 
\cite{balantekin, Takahashi3, Takahashi, Lunardini08}).
For example,
the initial progenitor mass dependence of the early neutrino burst
taking account of the MSW effect was
studied in \cite{Takahashi5},
and the Earth matter effects of the supernova neutrinos were
identified in  
\cite{Lunardini, Takahashi2, Takahashi4, Dighe03b}.
Moreover, neutrino spin-flavor conversion in supernova has been studied
considering numerically in
\cite{Ando03a, Ando03b, Ando03c} and 
analytically in \cite{Akhmedov03,  Ahriche03}.

Recently,
the effect of the shock wave on MSW effect of neutrino oscillations in
supernova  was
studied \cite{Schirato02, Lunardini03}. 
This effect 
appears as a decrease of the average energy of 
$\nu_e$ in the case of normal mass hierarchy (or $\bar{\nu}_e$ in the
case of inverted mass hierarchy)
\cite{Takahashi1, kawagoe2}.
Since the shock wave passes through H-resonance region, whose typical
resonance density is 
$\sim$ $10^3$ g/cm$^3$ in  a few seconds after core bounce, 
the adiabaticity of the resonance changes.
The time dependence of the neutrino events monitors the density profile
of the supernova 
\cite{Tomas, Fogli3, Fogli2, Takahashi1, TAUP2007}.
Therefore, using MSW effects embedded in the supernova neutrinos, we are
able to 
find the density profile of the supernova. 
However, in many previous studies
the influences of the shock wave
on the neutrino events were discussed schematically using
simplified and 
parameterized shock-wave profiles \cite{Fogli2}. 



There are many supernova simulations taking account of 
various explosion mechanisms. 
However, there are a few studies in which the neutrino
oscillations were calculated 
by using simulation results because  
almost all supernova simulations were performed in  only core region
although 
MSW effect occurs at outer envelope of the star.
In practice, it is
still very hard to carry out multidimensional simulation with neutrino
transport to 
proceed to the shock propagation from the core throughout the envelope.
Therefore, it is important for the studies of the neutrino oscillations
to calculate 
the shock propagation throughout a 
supernova not only in the core but also in the envelope.


In this paper,
we study how the shock wave propagation and the neutrino oscillation
parameters affect the supernova neutrinos.  
We calculate neutrino oscillations of the
supernova neutrinos by using supernova simulation result 
in adiabatic explosion model.
In our model, we calculate all processes of core collapse, bounce, and
shock propagation
continuously in a unified manner.
In order to clarify detailed dependence of the result on the neutrino
oscillation 
parameters and the shock wave,
we calculate survival probabilities, energy spectra, and the event
rates of 
three flavors of supernova neutrinos 
to be detected 
with SK and SNO.

In Sec. \ref{sec:level2}, we introduce our numerical method.
In Sec. \ref{sec:level3}, we show time evolution of the survival
probability and energy spectra of the supernova neutrinos.
We also study the event rates.
We analyze the dependence of these neutrino signals on neutrino
oscillation parameters and the shock effect.
We discuss the ratio of high- to low-energy neutrino events which will
be detected by SK and SNO. 
We discuss the difference of the expected supernova neutrinos using
both 
parameterized density and the density obtained by the simulations of
supernova explosion in
Sec. \ref{sec:level4}. 
We also discuss other effects of supernova explosions and supernova
neutrinos which should affect the time evolution of the neutrino signal. 
Finally, we conclude our paper in Sec. \ref{sec:level5}.

\section{\label{sec:level2}Numerical Method}

\subsection{\label{sec:level2A}Supernova Model}
In order to obtain the detailed density profile in a supernova explosion 
and to
examine the MSW effect on supernova neutrinos,
we use one dimensional simulation result of the supernova.
We model the supernova explosion using an implicit Lagrangian code
for general relativistic spherical hydrodynamics \cite{Yamada}.
The numerical tables of Shen's relativistic equation of
state (EOS) \cite{ShenEOS} 
and Timmes's EOS \cite{TimmesEOS} are adopted for the high and low
density matters in this code, respectively.
We adopt the presupernova model of a 15$M_{\odot}$ star
provided by Woosley and Weaver (WW95) \cite{Woosley} as the initial
condition. 
We use the distribution of the electron fraction $Y_e$ in WW95.
We perform calculations of all processes throughout the core collapse,
bounce, and shock propagation 
continuously by adiabatic collapse with fixed electron fraction to
follow the shock wave for a long time scale ($> 10$ s) and a wide
density range, consistently \cite{Adiabatic}.
We thus succeeded in the calculation of the propagation of the shock, which
is generated by the collapse of the iron core, until the shock wave
reaches the hydrogen rich envelope. %
Throughout one calculation, one finds that the density of the central
core reaches  
$\sim 10^{15}$ $\mathrm{g/cm}^3$ and the shock wave propagation for
beyond the 
hydrogen envelope ($\sim 1 $ $\mathrm{g/cm}^3$) is solved simultaneously
\cite{YAMADA, dthesis}. 
%
%

\begin{figure}
 \begin{center}
    \includegraphics[height=21pc]{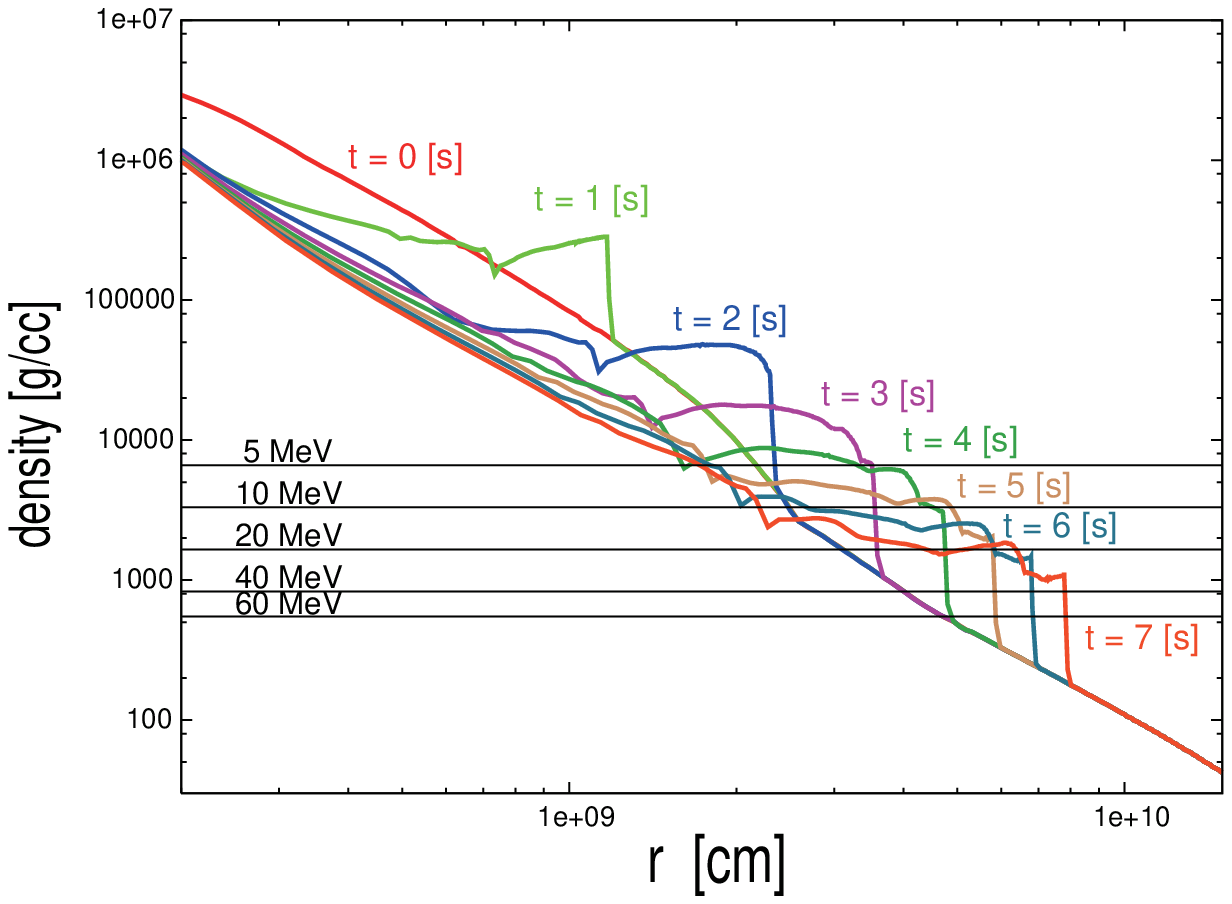}
     \caption[Density profile]
{\label{fig:ele-density} The density profiles as a function of radius for
  every 1 second from 0 to 7 seconds after the core bounce. The
  horizontal 
  lines show the density at the H-resonance point of each neutrino
  energy. }
 \end{center}
\end{figure}

Figure \ref{fig:ele-density} shows the calculated density profiles as
a function of 
radius for every 1 s from 0 to 7 s after the core bounce.
The shock wave reaches the oxygen-rich layer in 1 s and passes through
the helium layer in 10 s.
The horizontal lines show the density at the H-resonance of the
neutrino energy. 
The shock wave reaches the H-resonance in about 2 s.
The density behind the shock wave hardly decreases
because 
we neglected neutrino cooling of the proto-neutron star in the present study.
Therefore, 
the shock velocity is kept almost constant at
$9\times 10^8 $ cm/s in this model. 
When the neutrino cooling of the proto-neutron star is 
taken into consideration, 
the density and the pressure behind the shock would inevitably decrease.
As a result, the shock wave would be decelerated. 
We study
detailed analysis of the cooling effect of the proto-neutron star
elsewhere.

We assume the neutrino energy spectra with Fermi-Dirac distributions. 
Numerical simulations of supernova explosions which include neutrino
transport suggested that the resultant neutrino spectra at
low-to-intermediate energies do not exactly follow Fermi-Dirac
distributions.
However, since we are interested in the event number of neutrinos to be
detected in a water Cherenkov detector, the neutrino signals are
sensitive to high energy neutrinos because
the neutrino-induced 
reaction cross sections in the detector are
proportional to the neutrino energy. 
As large as the neutrino energies above $\sim 10$ MeV are concerned,
therefore, Fermi-Dirac distributions with finite chemical potentials may
be justified as one of the reasonable approximations to the results of
numerical simulations of the neutrino transport (\cite{Keil2003}; and
references therein).
Although several simulations suggest different neutrino temperatures
from one another, 
they all satisfy a common hierarchy 
$T_{\nu_e} < T_{\bar{\nu}_e} < T_{\nu_x} $.
The interactions between $\nu_e$ and neutrons freeze out at  lower
temperature and density than the interactions between $\bar{\nu}_e$ and
protons do inside the neutralized supernova core,  while the
interactions between $\nu_x $ ($=\nu_{\mu}$, $\bar{\nu}_{\mu}$,
$\nu_{\tau}$, and $\bar{\nu}_{\tau}$) and nucleons freeze out at  much
higher  temperature and density because $\nu_x$ interact  with matter
only through the neutral current.
In the present study,
the neutrino temperatures are set to be 
$T_{\nu_e} = 3.5$ MeV, $T_{\bar{\nu}_e} = 4$ MeV, and 
$T_{\nu_x} = 7$ MeV associated with
the neutrino chemical potentials 
$\mu_{\nu_e}=7.28$ MeV, $\mu_{\bar{\nu}_e}=10$ MeV, and
$\mu_{\nu_x}=0$ MeV.
These parameter values are well approximated to the time integrated
spectra of the  neutrinos obtained from a supernova simulation in
Livermore Group \cite{totani-sato-del}.
Previous studies of energy spectra of supernova neutrinos in Galactic
chemical evolution of $^{11}$B and $^{10}$B 
\cite{Yoshida2005, Yoshida2008} and the r-process nucleosynthesis 
\cite{Woosly1994, Yoshida2004} indicated that the neutrino temperatures
are constrained by the theoretical fit to the observed elemental
abundances of these nuclei : 
$T_{\nu_e}=2.5-3.5$ MeV, $T_{\bar{\nu}_e}=4-5$ MeV, and $T_{\nu_x}=4-7$
MeV, still satisfying hierarchy condition 
$T_{\nu_e}<T_{\bar{\nu}_e}<T_{\nu_x}$, where the uncertainty due to
finite chemical potentials \cite{Yoshida2005} are taken into account in
the inferred temperature values.
Our adopted values 
$T_{\nu_e}=3.5$ MeV, $T_{\bar{\nu}_e}=4$ MeV, and $T_{\nu_x}=7$ MeV are
in reasonable agreement with these constraints.
The energy spectrum of each flavor of  neutrinos is presumed to be
independent of time.
The total neutrino energy carried by neutrinos is set equal to 
$3 \times 10^{53}$
erg and the energy is equipartitioned to three-flavors of neutrinos and 
anti-neutrinos. 
We assume the exponential decay of the neutrino luminosity with 
a decay time of 3 s.
%
%

\subsection{\label{sec:level2B}Neutrino Oscillation}
In order to calculate neutrino oscillation effect on the neutrino
spectrum, we 
solve the time evolution of the neutrino  
wave function along the
density profile of our supernova model. 
The time evolution of the neutrino wave function is solved using the
equation 
\begin{eqnarray}
\imath \frac{d}{dt}
 \left( 
   \begin{array}{c}
     \nu_e \\ \nu_{\mu} \\ \nu_{\tau} 
   \end{array}
 \right)
=
\left\{
U
 \left(
   \begin{array}{ccc}
     0  & 0           & 0 \\
     0  & \Delta E_{12} & 0 \\
     0  & 0           & \Delta E_{13} 
   \end{array}
 \right)
U^{-1}
+ 
 \left(
   \begin{array}{ccc}
     \sqrt{2}G_{\mathrm{F}}n_e(r) & 0 & 0\\
     0  &  0  &  0  \\
     0  &  0  &  0 
   \end{array}
 \right)
\right\}
 \left( 
   \begin{array}{c}
     \nu_e \\ \nu_{\mu} \\ \nu_{\tau} 
   \end{array}
 \right), \label{eq:schredinger}
\end{eqnarray}
where $\Delta E_{ij}=\Delta m^2_{ij}/2E_{\nu}$,
$\Delta m^2_{ij} \equiv m^2_j - m^2_i$, $G_{F}$, $E_{\nu}$ and $n_e(r)$
are the energy difference between two mass eigenstates,
mass squared
differences, the Fermi constant, the  neutrino energy, and the 
electron number density.
In case of anti-neutrinos, the sign of $\sqrt{2}G_{F}n_e$
changes. 
$U$ is the  Cabibbo-Kobayashi-Masukawa (CKM) matrix
\begin{eqnarray}
\begin{array}{l}
U= 
 \left(
 \begin{array}{ccc}
           c_{12}c_{13} & s_{12}c_{13} & s_{13}e^{-i\delta }\\
         -s_{12}c_{23}-c_{12}s_{23}s_{13}e^{i\delta } & 
         c_{12}c_{23}-s_{12}s_{23}s_{13}e^{i \delta } &
         s_{23}c_{13} \\
         s_{12}s_{23}-c_{12}c_{23}s_{13}e^{i\delta }  &
         -c_{12}s_{23}-s_{12}c_{23}s_{13}e^{i \delta } &
         c_{23}c_{13}
 \end{array}
\right),
\end{array}
\end{eqnarray}
where $s_{ij}=\sin {\theta}_{ij}$, and $c_{ij}=\cos {\theta}_{ij}$
($i\neq j =1,2,3$) \cite{neutrino}.
We here put the CP violating phase $\delta $ equal to zero in the
CKM matrix.

The neutrino oscillation parameters are taken from the analysis of
various neutrino experiments \cite{Fukuda2001, sno}, except for $\theta
_{13}$ 
\cite{Parameter}, as 
\begin{eqnarray}
\sin ^2 2 \theta _{12}=0.84, \;\;\;\;\;
\sin ^2 2 \theta _{23}=1.00,
\label{eq:theta12}
\end{eqnarray}
and
\begin{eqnarray}
\Delta m^2_{12}=8.1 \times 10^{-5} \mathrm{eV}^2, \; \;\;\;\;
| \Delta m^2_{13} | =2.2 \times 10^{-3} \mathrm{eV}^2.
\end{eqnarray}
We calculate the neutrino survival probabilities for both  normal 
($\Delta m^2_{13} > 0$) and inverted ($\Delta m^2_{13} < 0$) mass
hierarchies for four cases of $\sin^2 2 \theta_{13}$ value: 
$\sin ^2 2\theta _{13}=10^{-2}$, $10^{-3}$, $10^{-4}$  and  $10^{-5}$.

The energy spectra of the neutrinos passed through the exploding
supernova,  
$\phi ^{\mathrm{SN}}_{\nu}(E_{\nu})$,
are obtained from 
the spectra of the neutrino emitted from the neutrino sphere, 
$\phi ^{\mathrm{org}}_{\nu}(E_{\nu})$, multiplied by the survival
probability 
in accordance with the following  equation,
%
%
\begin{widetext}
\begin{eqnarray}
\left( 
 \begin{array}{c}
  \phi ^{\mathrm{SN}}_{\nu_e}(E_{\nu}) \\
  \phi ^{\mathrm{SN}}_{\bar{\nu}_e}(E_{\nu}) \\
  \phi ^{\mathrm{SN}}_{\nu_x}(E_{\nu})
 \end{array}  
\right)
= 
\left(
 \begin{array}{ccc}
  P(E_{\nu}) & 0 & 1-P(E_{\nu}) \\
  0 & \bar{P}(E_{\nu})  & 1-\bar{P}(E_{\nu}) \\
  1-P(E_{\nu}) & 1-\bar{P}(E_{\nu}) &
                  2+P(E_{\nu})+\bar{P}(E_{\nu})
 \end{array}
\right)
\left(
 \begin{array}{c}
  \phi ^{\mathrm{org}}_{\nu_e}(E_{\nu}) \\
  \phi ^{\mathrm{org}}_{\bar{\nu}_e}(E_{\nu}) \\
  \phi ^{\mathrm{org}}_{\nu_x}(E_{\nu})
 \end{array}  
\right), \label{eq:spctra}
\end{eqnarray}
\end{widetext}
where 
$\phi_{\nu_x} \equiv
\frac{1}{4} 
(\phi_{\nu_{\mu}}+\phi_{\nu_{\tau}}+\phi_{\bar{\nu}_{\mu}}
+\phi_{\bar{\nu}_{\tau}})$, 
and
{\it P} and $\bar{P}$ are survival probabilities of $\nu_e$ and
$\bar{\nu}_e$, respectively. 

Neutrinos change largely their flavors at resonances of neutrino
oscillations. 
There are two resonances for supernova neutrinos (e.g., \cite{dighe}). 
The resonance points of higher and lower electron number densities
correspond to H-resonance and L-resonance, respectively.
The electron number density at the resonance point is written as 
\begin{eqnarray}
n_{e, res}\equiv \frac{1}{ 2 \sqrt{2} G_F}
                \frac{\Delta m^2}{E_{\nu}}
                \cos 2 \theta,
\label{eq:resonance-density}
\end{eqnarray}
where $\Delta m^2$ and $\theta $ correspond to $|\Delta m^2_{13}|$ and 
$\theta _{13}$ at
H-resonance and to $\Delta m^2_{12}$ and $\theta _{12}$ at
L-resonance, respectively.
The flavor change at a resonance strongly
depends on adiabaticity of the resonance.
When the resonance is adiabatic, large flavor change occurs.
The neutrino adiabaticity is estimated by $\gamma$,
\begin{eqnarray}
\gamma \equiv 
\frac{\Delta m^2}{2E_{\nu}}\frac{\sin^2 2\theta}{\cos  2 \theta}
\frac{n_{e, res}}{|\frac{dn_e}{dr}|}.
\label{eq:gamma}
\end{eqnarray}
If $\gamma$ is larger than 1,
the resonance becomes adiabatic, but if it is small,
resonance becomes non-adiabatic.
The  change of adiabaticity mainly occurs at
H-resonance owing to  the shock propagation. 
When the shock wave reaches the H-resonance, the density gradient becomes
large and the value of $\gamma $ 
decreases.
This leads to a change of survival probability of neutrinos.
Since the resonance point is inversely proportional to the neutrino
energy, 
this effect appears from low-energy side and moves toward high-energy
side according to the shock wave propagation.
As shock wave propagates outward, the density at the
shock front decreases and the resonance condition is satisfied for
higher energy neutrinos.


\subsection{\label{sec:level2c}Detection of Supernova Neutrinos}
The expected event number of the neutrinos in a water
Cherenkov detector can
be expressed as
\begin{eqnarray}
\frac{d^2 N}{dE_e dt}=N_{tar} \, \eta_{(E_e)} \,
  \frac{1}{4 \pi d^2} \, \frac{d^2 N_{\nu}}{dE_{\nu }dt} \,
   \sigma(E_{\nu}) \,  \frac{dE_{\nu}}{dE_{e}}
\label{eq:ivent},
\end{eqnarray}
where $N$ is the detection number of neutrinos, 
$E_e$ is the energy of electron or positron, 
$E_{\nu}$ is the energy of neutrino, 
$N_{tar}$ is the target number, 
$\eta_{(E_e)}$ is the efficiency of the detector, 
$d$ is the distance from the supernova, 
$\frac{d^2N_{\nu}}{dE_{\nu}dt}$ is the neutrino
spectrum,  and
$\sigma (E_{\nu })$ is the cross section as a function of the
neutrino energy \cite{neutrino}.
We assumed a supernova at the center of
Milky Way ($d$=10 kpc), and 
we neglected the Earth effect \cite{keitaro}.
We assume the detection at  SK and SNO.

The mass of SK is 32000 ton and the solvent is
pure water. 
SK mainly detects the $\bar{\nu}_e$ events with the
reaction $\bar{\nu}_e + p \rightarrow e^+ + n$. 
The event number is obtained by integrating over the angular
distribution of the events. 
As for the efficiency of the detector, we assumed as follows for
simplicity: 
$\eta _{(E_e)}=0$ for $E_e < 7$ MeV and
$\eta _{(E_e)}=1$ for $E_e \geq 7 $ MeV in accordance with
SK Phase II \cite{PhaseII}.
The finite energy resolution of the detector was neglected here.
%
%
The $\bar{\nu}_e$ and $\nu_e$ events are evaluated
taking account of the following four reactions,
\begin{eqnarray}
\bar{\nu_e}+ p & \rightarrow & e^+ + n \;,
\label{eq:eb+p}\\
\nu_e+ e^- & \rightarrow & \nu_e+ e^-,
\label{eq:e+e}
\\
\bar{\nu_e}+ e^- & \rightarrow & \bar{\nu_e}+ e^-,
\\
\nu_x+ e^- & \rightarrow & \nu_x+ e^-.
\end{eqnarray}
The energies of positron and electron are evaluated as
$E_{e^{+}} = E_{\nu}+m_p-m_n-m_{e{+}}$ for  
$\bar{\nu_e} + p \rightarrow n + e^{+}$
  and $E_e = E_{\nu}- \frac{m_e}{2}$ for 
$\nu + e^- \rightarrow \nu + e^-$, 
respectively.
The cross sections of the above four reactions are adopted from 
\cite{neutrino}.


The mass of SNO is 1000 ton and the solvent is
heavy water. 
SNO detects not only $\bar{\nu}_e$ events but also $\nu_e$  events with  
charged and neutral current reactions. 
As for the efficiency of the detector, we assumed as follows for
simplicity: 
$\eta _{(E_e)}=0$ for $E_e < 5$ MeV and
$\eta _{(E_e)}=1$ for $E_e \geq 5 $ MeV \cite{Bahcall2000}.
We evaluate the $\bar{\nu}_e$ and $\nu_e$ event numbers using the four
reactions, 
\begin{eqnarray}
\nu_e + d & \rightarrow & p+p+e^- \\
\bar{\nu_e}+d & \rightarrow & n+n+e^+  
\label{eq:sno}\\
\nu
+ d & \rightarrow & p+n+\nu \\
\nu + e^- & \rightarrow & \nu + e^- .
\end{eqnarray}

The main reactions of $\bar{\nu}_e$ and 
$\nu_e$ are the first and second reactions, respectively
\cite{Ying1989}.
The cross sections of these reactions are adopted from
\cite{nakamura2001}.  
The cross sections become larger as 
the neutrino energy increases.
Although
the SNO experiment has already ended,
we also evaluate the neutrino events by 
SNO 
because the detection
efficiency of $\nu_e$ is high and similar in next generation detector like
SNO.

\section{\label{sec:level3}Result and Discussions}
\subsection{Survival probability}

\begin{figure}
 \begin{center}
   \resizebox{165mm}{!}{\includegraphics{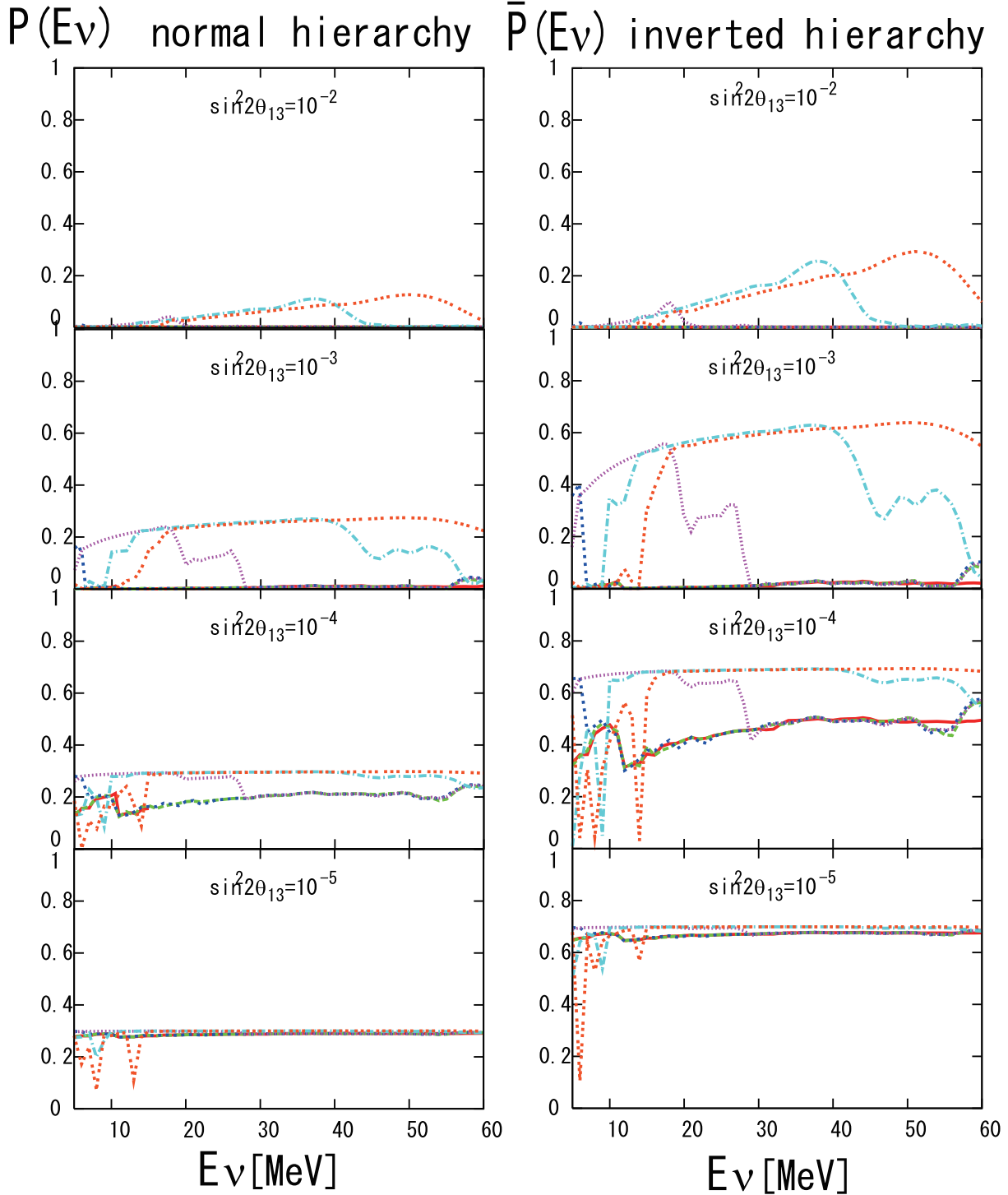}}
 \end{center}
\caption{\label{fig:survival-normal}Calculated results of the survival
 probabilities, $P(E_{\nu})$ and $\bar{P}(E_{\nu})$, 
in every 1 second.
Left panels are the survival probabilities of $\nu_e$,
$P(E_{\nu})$, in normal
 hierarchy, and right panels are the survival probabilities of
 $\bar{\nu }_e$, $\bar{P}(E_{\nu})$, in inverted hierarchy.
Red, green, blue, purple, sky blue and orange lines are t = 0, 1, 2, 3,
4 and 5 sec after core bounce. 
The value of $\sin ^2 2 \theta_{13}$ is 
$10^{-2}$, $10^{-3}$, $10^{-4}$ and $10^{-5}$
 from top to bottom. }
\end{figure}

We calculate the survival probabilities of $\nu_e$ and $\bar{\nu}_e$
from the supernova,
$P(E_{\nu})$ and $\bar{P}(E_{\nu})$.
Figure \ref{fig:survival-normal} shows the calculated results. 
Left panels of Figure \ref{fig:survival-normal} are $P(E_{\nu})$
in the case of normal hierarchy 
in every 1 s, and right panels of Figure
\ref{fig:survival-normal}
are $\bar{P}(E_{\nu})$ in the case of inverted hierarchy. 
Red, green, blue, purple, sky blue and orange lines correspond to the
results in  
$t$ = 0, 1, 2, 3,
4 and 5 s after core bounce. 
We showed the results for the values of 
$\sin ^2 2 \theta _{13}=10^{-2}$, $10^{-3}$, $10^{-4}$ and $10^{-5}$
 from top to bottom.
In the case of  $\sin ^2 2 \theta_{13} = 10^{-2}$ and $10^{-3}$,
$P(E_{\nu})$ and $\bar{P}(E_{\nu})$ are about 0
when shock wave does not reach H-resonance
($t$ = 0 $-$ 2 s).
However,  in later time ($t$ = 2 $-$ 5 s),
$P(E_{\nu})$ and $\bar{P}(E_{\nu})$ become finite. 
These effects appear from low-energy side and gradually shift toward
high-energy  
side as the time passes by.
In the case of $\sin^2 2\theta_{13} = 10^{-5}$, 
$P(E_{\nu})$ and $\bar{P}(E_{\nu})$ hardly change with time,
and are about 0.3 and 0.7, respectively. 
In the case of $\sin ^2 2\theta_{13} = 10^{-4}$,
$P(E_{\nu})$ and $\bar{P}(E_{\nu})$ are  about 0.2 and 0.4 in $t = 0 - 2$
s, respectively.
These survival probabilities indicate intermediate adiabaticity between 
the cases for large $\theta_{13}$ and small $\theta_{13}$.
At the later times they become large as shown in the large $\theta_{13}$ 
case. 

We can understand these behaviors considering the shock wave
propagation.  
The adiabaticity of H-resonance is estimated by $\gamma$ in 
Eq. (\ref{eq:gamma}). 
When the shock wave reaches H-resonance region, the adiabaticity
changes. 
If $\sin^2 2 \theta_{13}$ is large and there
is not the shock wave at the resonance region,
$\gamma $ is larger than 1, and the resonance is adiabatic.
As a result, the survival probabilities are 0.
When the shock wave reaches a resonance, large density gradient reduces
$\gamma$ and the resonance becomes non-adiabatic. 
Thus, the survival probabilities become finite during the shock passage
in the resonance region.
In contrast, if $\sin ^2 2\theta_{13}$ is very small,
$\gamma $ is smaller than 1, and the resonance is non-adiabatic.
Consequently, the survival probabilities are finite, and
 the influence of the shock wave hardly appears to the
survival probability,
since H-resonance totally in non-adiabatic regardless the existence
of the shock wave.
%
In the case of $\sin^2 2 \theta_{13} =10^{-2}$, however, the influence
of the shock wave is not clearly seen though the $\sin^2 2\theta_{13}$
is large. 
This is because that $\sin ^2 2 \theta_{13}$ is very large and $\gamma$
is larger than 1 even when $|\frac{1}{n_e}\frac{dn_e}{dr}|$ becomes large.

\begin{figure}[p]
    \includegraphics[height=45pc]
    {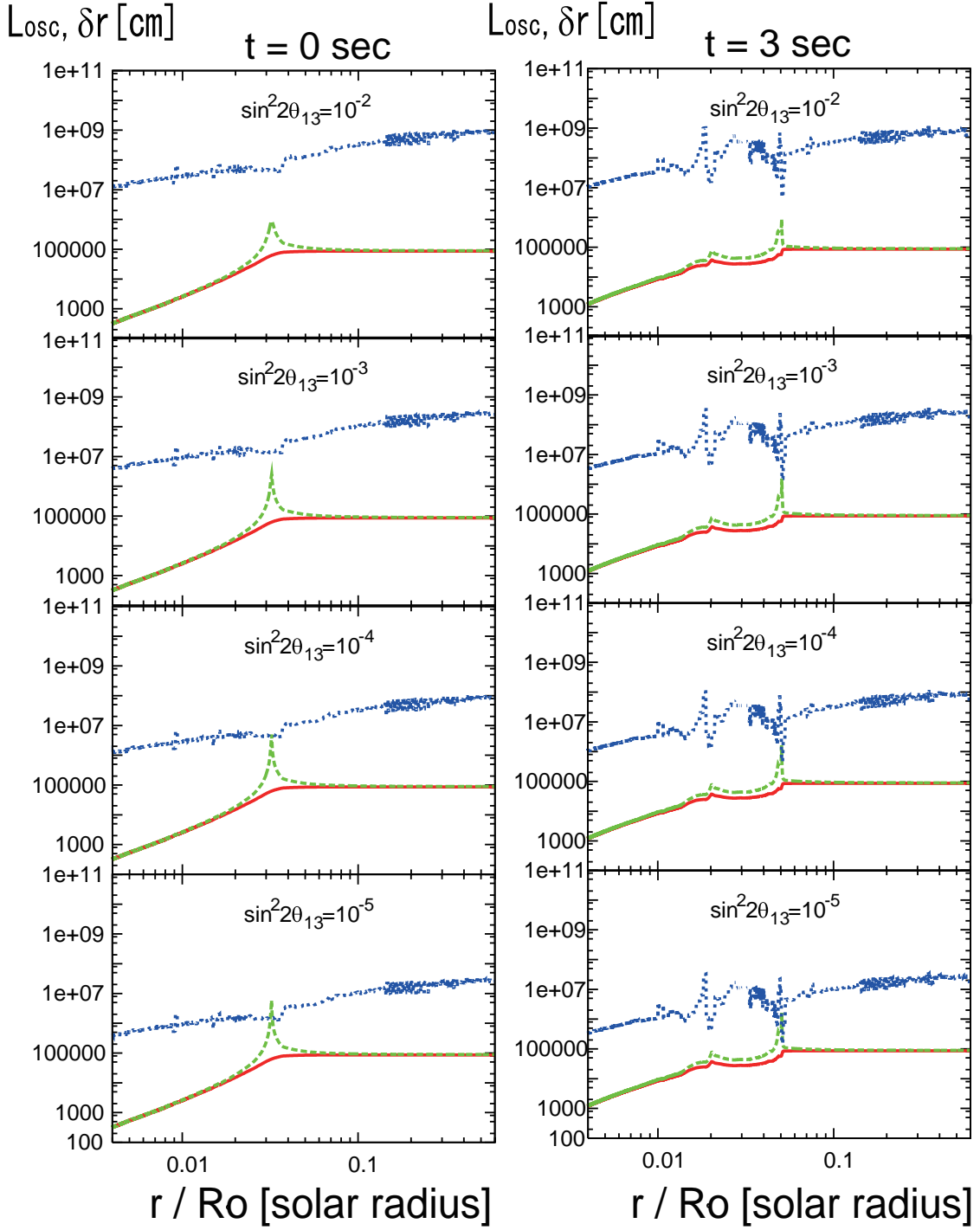}
 \caption[The oscillation in inverted hierarchy]
{\label{fig:inv-oscillation-length}
The oscillation length,
$L_{\mathrm{osc}}=\frac{2E}{\Delta \tilde{m}^2 \sin 2 \tilde{\theta}}$,
and the level crossing length,
$\delta r = \frac{\sin 2 \theta}{\cos 2 \theta} 
|\frac{1}{n_e}\frac{dn_e}{dr}|^{-1}$, 
of 
neutrinos for 5 MeV
 in the case of inverted mass hierarchy as a function of a radius $r$ in
 units of solar radius 
$R_{\odot} = 6.96 \times 10^{10}$ cm.
The left panels are at $t$ = 0 sec, 
and the right panels are at $t$ = 3 sec, respectively. 
We calculate by the value of $\sin ^2 2\theta_{13}=10^{-2}$, $10^{-3}$,
$10^{-4}$ and $10^{-5}$ from top to bottom. 
Red and green lines are the oscillation length for 13 mass eigenstate
 for neutrinos and anti-neutrinos,
and blue line are the level crossing length.}
\end{figure}

$\gamma$ can be rewritten by the ratio of the
length characterizing the neutrino oscillations
$L_{\mathrm{osc}}=\frac{2E_{\nu}}{\Delta \tilde{m}^2 \sin 2
\tilde{\theta}}$,
where $\Delta \tilde{m}^2 = $
$\{(2\sqrt{2}G_F E_{\nu } n_e(r) / \Delta m^2 - \cos2 \theta)^2+\sin^2
2\theta\}^{1/2}\Delta m^2$,
to the length of the level crossing region, i.e. the level crossing
length
$\delta r = \frac{\sin 2 \theta}{\cos 2 \theta} 
|\frac{1}{n_e}\frac{dn_e}{dr}|^{-1}$ at resonance. 
Therefore, 
the adiabaticity of the neutrinos can be evaluated by 
$L_{\mathrm{osc}}$ and $\delta r$ \cite{neutrino}.
The resonance is adiabatic 
when the level crossing length is larger than the oscillation length,
$L_{\mathrm{osc}}\ll \delta r$.
Figure \ref{fig:inv-oscillation-length} shows the level crossing region
and the
oscillation length of neutrinos ($E_{\nu}$ = 5 MeV) in the case of inverted
mass 
hierarchy as a function of the radius.
The left panels of Figure \ref{fig:inv-oscillation-length} are at $t$ = 0
s, 
and the right panels are at $t$ = 3 s, respectively. 
Shown are the results for
$\sin ^2 2\theta_{13}=10^{-2}$, $10^{-3}$,
$10^{-4}$ and $10^{-5}$ from top to bottom. 
Red and green lines display the oscillation lengths of $\nu_e$ and
$\bar{\nu}_e$, 
and blue line shows the level crossing length using $\theta_{13}$.
%
%
The oscillation length of
$\bar{\nu}_e$ becomes extremely large on 
the resonance, and its behavior 
depends on $\sin ^2 2\theta_{13}$ below or above the resonance strongly.
On the other hand, the oscillation length of $\nu_e$ 
changes rather
gently as a function of radius, and strong dependence on 
$\sin ^2 2\theta_{13}$ is  not clearly seen.
As $\sin ^2 2\theta_{13}$ becomes
smaller or the shock wave reaches the resonance, the level crossing length
becomes shorter.

In the left panels ($t = 0$ s), the level crossing length at  
$\sin ^2 2\theta_{13}=10^{-2}$ and $10^{-3}$ are larger than the
oscillation length at  all region.
Therefore, 
the resonance is adiabatic.
On the other hand, 
in the case of 
$\sin^2 2 \theta_{13}=10^{-4}$ and $10^{-5}$, 
the level crossing length at the resonance
is almost 
same as or slightly smaller than
the oscillation length of $\bar{\nu}_e$.
Therefore, 
the resonance is non-adiabatic.
In the right panels ($t = 3$ s) 
when the shock wave reaches the H-resonance,
the level crossing length, $\delta r$,
is of the same order
or smaller than the oscillation length $L_{\mathrm{osc}}$ for all cases
of the mixing angle,
expect for 
$\sin ^2 2 \theta_{13}=10^{-2}$.
This satisfies non-adiabatic condition
$\gamma < 1$ as discussed above. 
We  understand, therefore,  that 
the resonance becomes
non-adiabatic by the effect of the shock wave.
%
%
%
%
In Figure \ref{fig:survival-normal}, the influence of the shock wave
appears about 2 s after core bounce.
This result is consistent with our simulation 
(see Figure \ref{fig:ele-density}).

In the case of normal hierarchy, the survival probability of $\bar{\nu}_e$
does not change much because there is no resonance in the $\bar{\nu}$
sector. 
Therefore, the survival probability of $\bar{\nu}_e$ in normal hierarchy
is always $\sim $ 0.7 regardless of the value of $\sin^2 2 \theta_{13}$
or independently of the shock wave.

We note that there is 
L-resonance in $\nu$ sector even in the case of inverted hierarchy.
However, the value of $\theta_{12}$ which is   related to L-resonance is
very large (see Eq. (\ref{eq:theta12})), and the level crossing length of our
simulation  is not as small as
the oscillation length at the resonance.
Therefore, $\gamma $ at the L-resonance is larger than 1.
As a result, the survival probability of $\nu_e$ in inverted
hierarchy does not  change drastically, and stays always $\sim$ 0.3.

\subsection{Supernova neutrino spectrum}

\begin{figure}
 \begin{center}
   \resizebox{140mm}{!}{\includegraphics{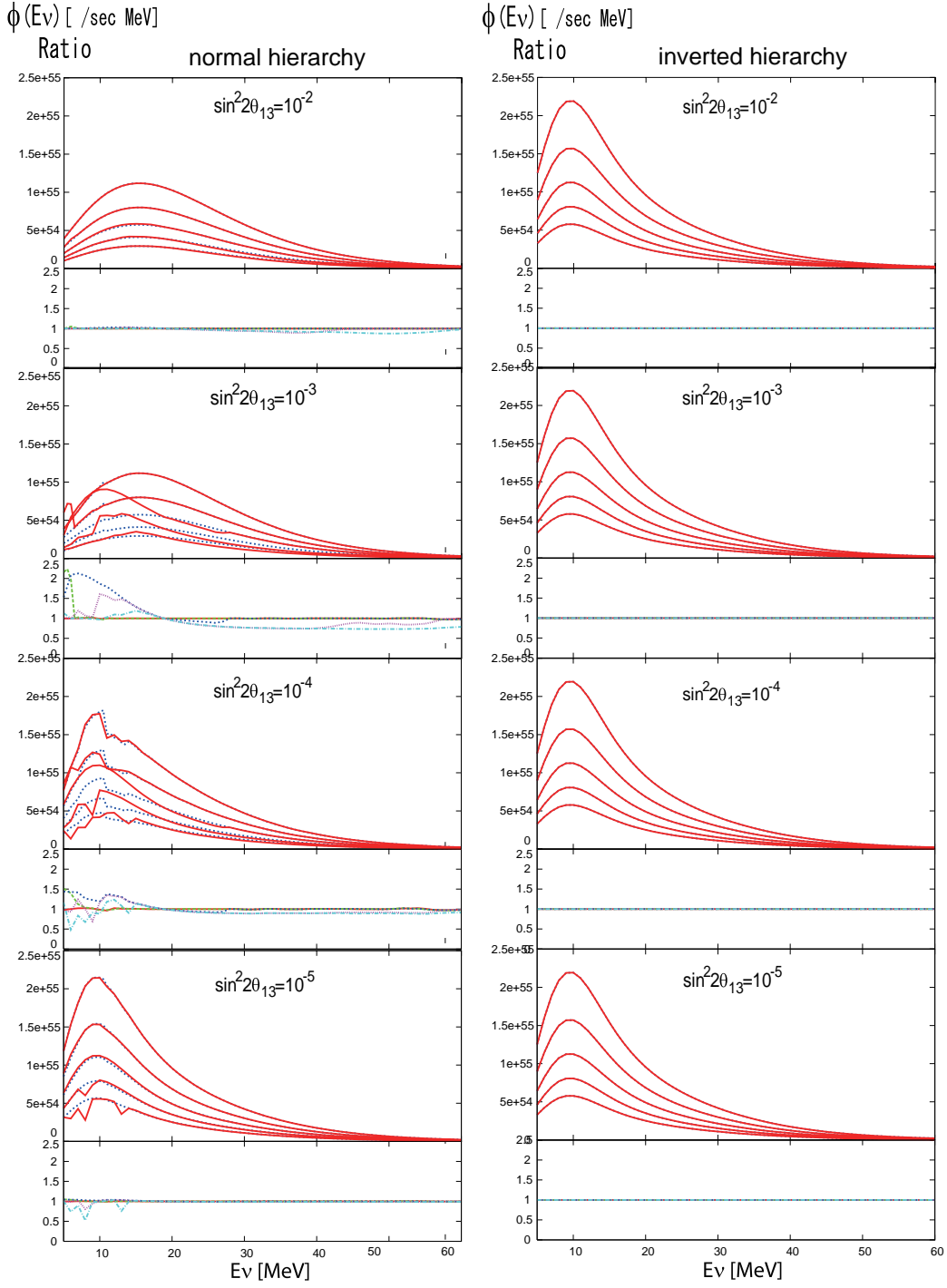}}
 \end{center}
 \caption{\label{fig:spctr-e}
Upper part of each panels are $\nu_e$ spectra 
$\phi (\nu_e)$ in every 1 second.
Solid (red) and dotted (blue) lines are the calculated results with
and without shock wave, respectively.
Lower part of each panel is the ratio of the spectra with shock to
 without shock,
$\phi_{with}(E_{\nu})/\phi_{without}(E_{\nu})$,
at $t = $ 1(red), 2(green), 3(blue), 4(purple) and 5(sky
 blue) sec.
The value of
 $\sin ^2 2\theta _{13}$ is $10^{-2}$, $10^{-3}$, $10^{-4}$ and
 $10^{-5}$ from top to bottom.
Left and right panels are in normal and inverted hierarchy, respectively.
}
\end{figure}

\begin{figure}
 \begin{center}
   \resizebox{140mm}{!}{\includegraphics{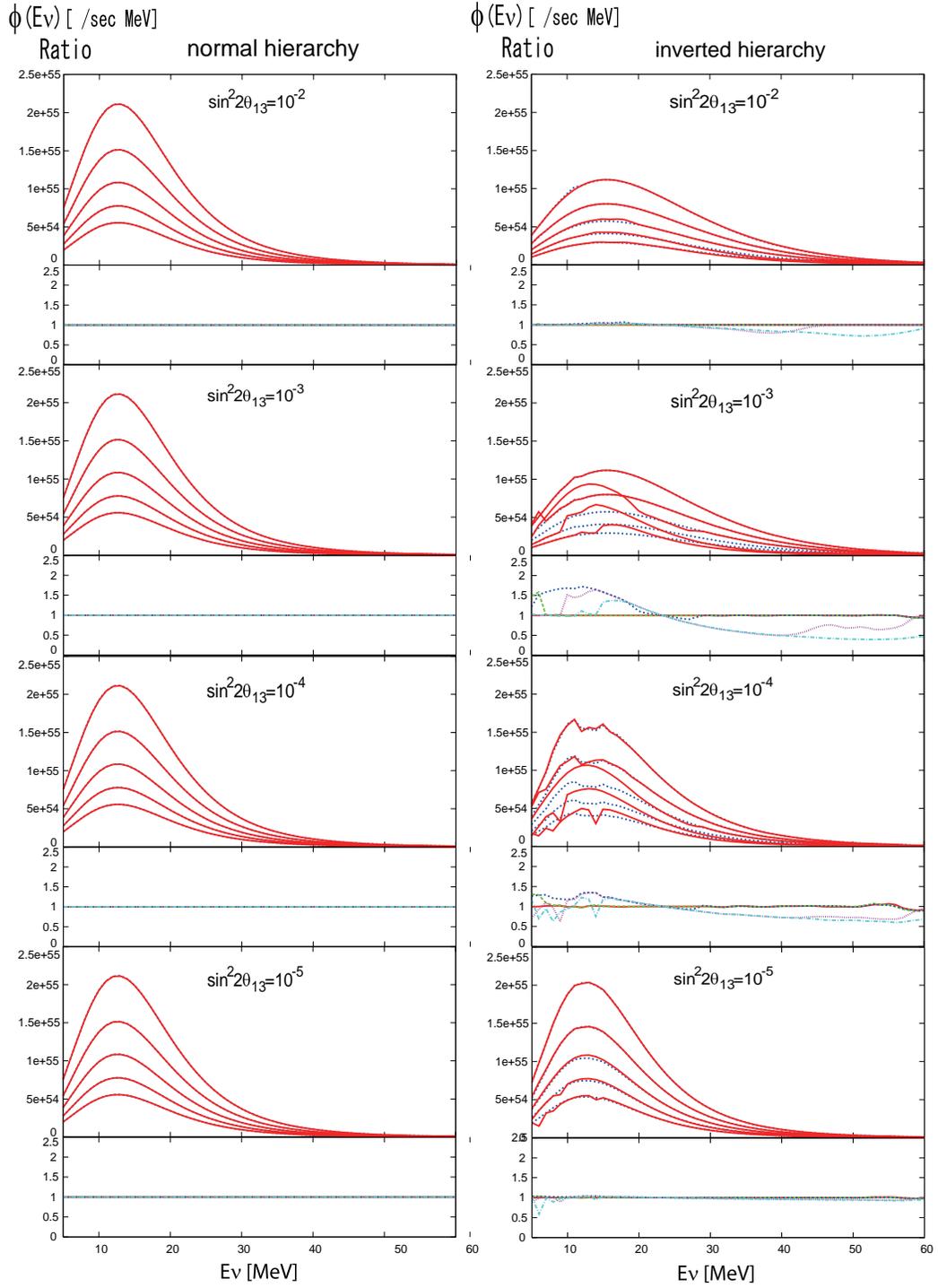}}
 \end{center}
 \caption{\label{fig:spctr-eb}
The same as Figure \ref{fig:spctr-e}, but for $\bar{\nu}_e$ spectra.
}
\end{figure}

We calculate the supernova neutrino spectra using the survival
probabilities.
Figure \ref{fig:spctr-e} shows the spectra of $\nu_e$.
Left panels of Figure \ref{fig:spctr-e} show the spectra in the case of
normal 
hierarchy, and right panels show the results of inverted hierarchy.
Figure \ref{fig:spctr-eb} is same as Figure \ref{fig:spctr-e} but for
$\bar{\nu}_e$ spectra.
Red solid  and blue dotted  lines are 
the spectra with and without shock wave, respectively.
In order to clearly observe the shock wave effects, we display these
ratio, $\phi_{\mathrm{with}}/ \phi_{\mathrm{without}}$, of the spectra
with to without shock
in lower part of each panel. 

We see clearly the shock wave effects in the $\nu_e$ spectra in Figure
\ref{fig:spctr-e} in normal hierarchy.
In the case of  
$\sin^2 2 \theta _{13} = 10^{-3}$ and $10^{-4}$,
an enhancement in low energy component of
the neutrino spectra is seen when the shock wave reaches the
H-resonance.
At later times, 
the influence of the shock wave on the spectra moves from the low-energy
side to the high-energy side.
The effect at later times is seen as a reduction of the high energy
component of the neutrino spectrum.
The influence of the shock wave in the case of 
$\sin^2 2\theta_{13} = 10^{-2}$
does not clearly appear in spectra
for the same reason as discussed in the previous section.
In the case of $\sin ^2 2 \theta_{13}= 10^{-5}$, 
the neutrino spectra 
with shock  are not different from the spectra without shock. 
H-resonance is non-adiabatic even without the shock wave.
We do not see any shock effects on the $\bar{\nu}_e$ spectra in inverted
hierarchy.

From Figure \ref{fig:spctr-eb}, the effect of shock wave appears in 
the spectra of $\bar{\nu}_e$ 
in the case of inverted hierarchy.
The dependence of the $\bar{\nu}_e$ spectra on $\sin ^2 2 \theta_{13}$ 
is almost the same as the dependence of the $\nu_e$ spectra in normal
hierarchy. 

The spectrum of $\nu_e$ at the surface  of the supernova is written from
Eq.(\ref{eq:spctra}) as
\begin{eqnarray}
\phi ^{\mathrm{SN}}_{\nu_e}(E_{\nu})
& =& P(E_{\nu}) \{\phi^{\mathrm{org}} _{\nu_e}(E_{\nu})
        -\phi^{\mathrm{org}} _{\nu_x}(E_{\nu})\}
         +\phi^{\mathrm{org}} _{\nu_x}(E_{\nu}).
\end{eqnarray}
The initial spectra 
are $\phi^{\mathrm{org}}_{\nu_e} > \phi^{\mathrm{org}}_{\nu_x}$ 
in low energy side, and 
($\phi^{\mathrm{org}} _{\nu_e} -\phi^{\mathrm{org}} _{\nu_x}$) is
positive. 
Therefore, $\phi ^{\mathrm{obs}}_{\nu_e}$ increases when $P$ is not 0.
On the other hand, the initial
spectra are $\phi^{\mathrm{org}}_{\nu_e} < \phi^{\mathrm{org}}_{\nu_x}$
in high energy side, and   
($\phi^{\mathrm{org}} _{\nu_e} -\phi^{\mathrm{org}} _{\nu_x}$) is
negative.
Therefore, $\phi ^{\mathrm{obs}}_{\nu_e}$ decreases when $P$ is not 0.
These increase and decrease appear in  the observed neutrino spectra as
the result from the  
influence of the shock wave.
In the lower part of each panel, we can see this effect clearly.
We note that the energy that satisfies  
$\phi^{\mathrm{org}} _{\nu_e} =\phi^{\mathrm{org}} _{\nu_x}$ is 
about 18.7 MeV, and the energy that satisfies 
$\phi^{\mathrm{org}} _{\bar{\nu}_e} =\phi^{\mathrm{org}} _{\nu_x}$ is 
about 23.3 MeV.

\subsection{Event rate of supernova neutrino}\label{sec:eventrate}

\begin{figure}
 \begin{center}
  \begin{tabular}{cc}
   \resizebox{92mm}{78mm}{\includegraphics{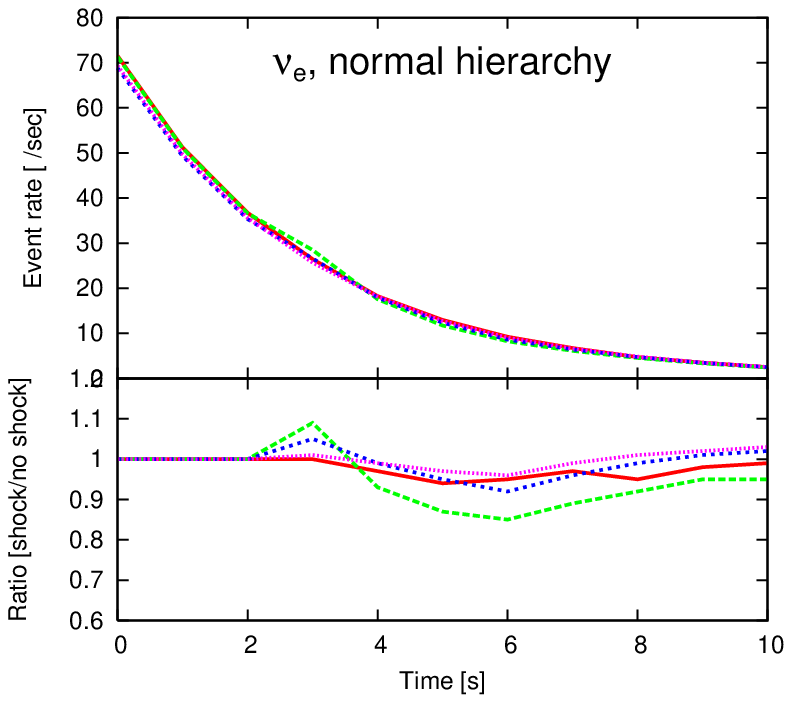}}&
   \resizebox{92mm}{78mm}{\includegraphics{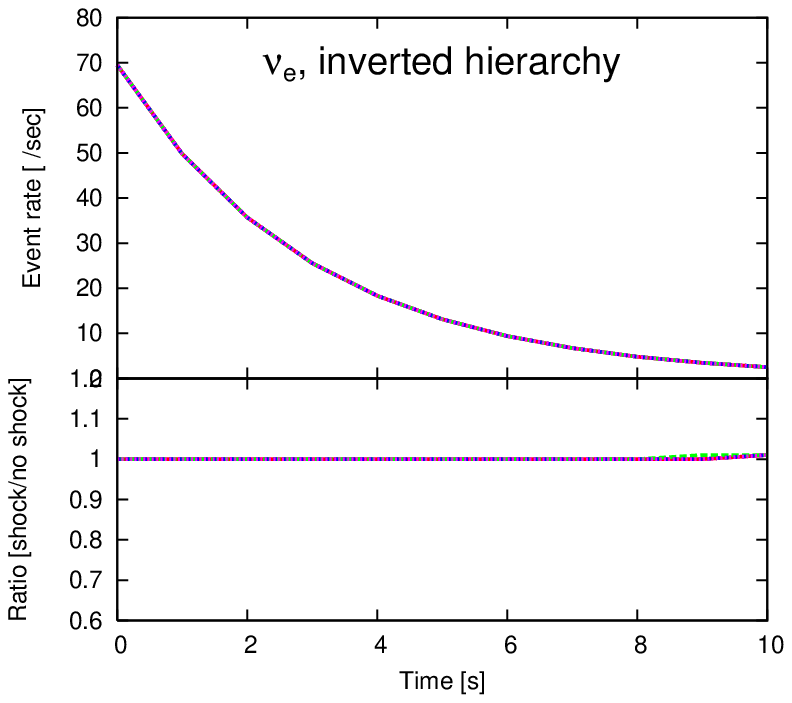}}
  \end{tabular}
 \end{center}
 \caption[The event e]
{\label{fig:eventrate-e}
Upper part of each panels shows the expected energy-integrated event
 rates of $\nu_e$  with
 SK as a function of time,
and lower part shows the ratio of these event rates with shock to those
 without 
 shock for various mixing angles $\sin ^2 2\theta_{13} = 10^{-2}$(red),
 $10^{-3}$(green), $10^{-4}$(glue) and $10^{-5}$(purple).
Left and right panels are in normal and inverted hierarchy,
 respectively. 
}
\end{figure}

We discuss here the predicted event rates of the supernova neutrinos to
be 
detected with SK in the manner described in section \ref{sec:level2c}.  
The upper parts of each panel of Figure \ref{fig:eventrate-e} show the 
expected energy-integrated event rates of $\nu_e$
\begin{eqnarray}
\frac{dN}{dt} \equiv \int_{E_{\mathrm{th}}}^{\infty}
\frac{d^2 N}{d E_e dt}d E_e,
\label{eq:integral-event}
\end{eqnarray}
where the integrand in right hand side   
is the expected event number of 
neutrinos defined by Eq. (\ref{eq:ivent}), and $E_{\mathrm{th}}$ are the 
threshold energies 
which are equal to 7 MeV and 5 MeV for SK and SNO
detectors, respectively, as explained in section \ref{sec:level2c}.
Left and right panels of Figure \ref{fig:eventrate-e} are for the cases
of 
normal and inverted  hierarchies, respectively.
The lower part of each panel 
shows the
ratio of  the event rates 
with and without shock.
We calculate for the four cases of the mixing angles: 
$\sin ^2 2\theta_{13} =10^{-2}$ (red), $10^{-3}$ (green), $10^{-4}$
(blue) and $10^{-5}$ (purple).

In the case of normal hierarchy (left panel of Figure
\ref{fig:eventrate-e}), 
the event rate of $\nu_e$ of each parameter is not separated clearly 
from one another.
The event rate is commonly about 70 at $t=0$ s and decreases to about 3
at $t=10$ s. 
%
On the other hand, the ratio of the event rates of 
 $\nu_e $ with and without shock 
shows some different  $\theta_{13}$ dependence (left lower panel of
Figure \ref{fig:eventrate-e}).
After 2 s, we can see slight enhancement of this ratio.
We find from this change that  
the shock front does not reach H-resonance before $\sim 2$ s. 
Once the shock wave reaches H-resonance region
the event rate increases from 
the low-energy neutrinos because of energy dependence of the resonance
density Eq. (\ref{eq:resonance-density}).
This ratio then decreases after 3 s because the event rate of
the high-energy neutrinos decreases when the
shock wave propagates through the resonance. 
The obtained time profile thus  
depends on the
$\theta_{13}$ values. 
The effect of shock propagation 
appears most clearly in the case of 
$\sin ^2 2 \theta_{13} = 10^{-3}$ (green).
However, the decrease in the ratio of the event rates is by at most 15
\%
%
in the case of small $\theta_{13}$.
It is smaller than the case for
$\sin ^2 2\theta_{13}=10^{-3}$
at any time for the other $\theta_{13}$ values.
%

In the case of the inverted mass hierarchy (right panel of Figure
\ref{fig:eventrate-e}),  
the time variation of the event rate of $\nu_e$ does not depend on
$\theta_{13}$.
Shock wave effect is not clearly seen (lower part of right panel of Figure
\ref{fig:eventrate-e}).
The adiabaticity of $\nu_e$ is not influenced by the 
shock wave
because $\nu_e$ is not related to H-resonance in the case of inverted
hierarchy.
Therefore, the spectra of $\nu_e$ do not change 
for any values of $\theta_{13}$. 

\begin{figure}[b]
 \begin{center}
  \begin{tabular}{cc}
   \resizebox{92mm}{78mm}{\includegraphics{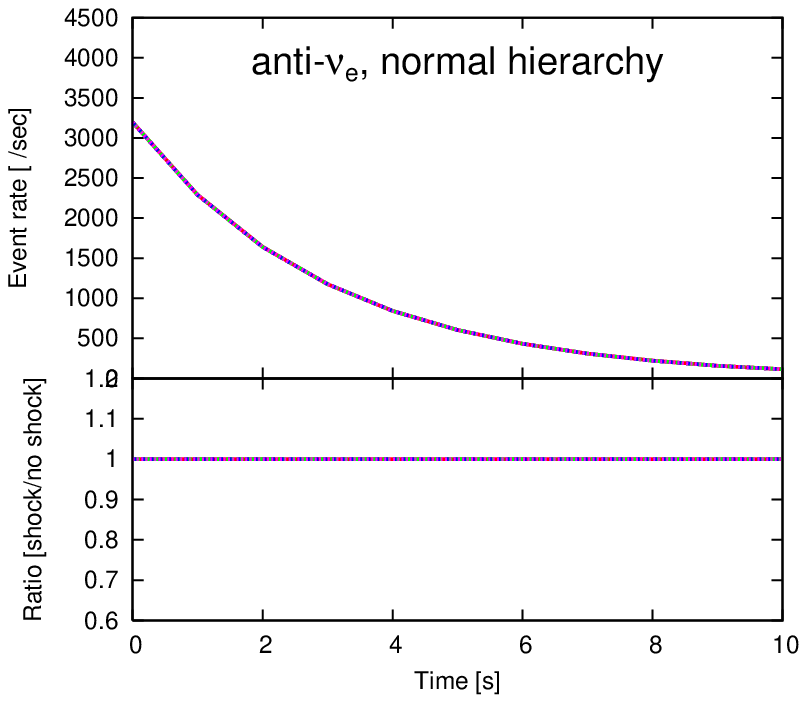}}&
   \resizebox{92mm}{78mm}{\includegraphics{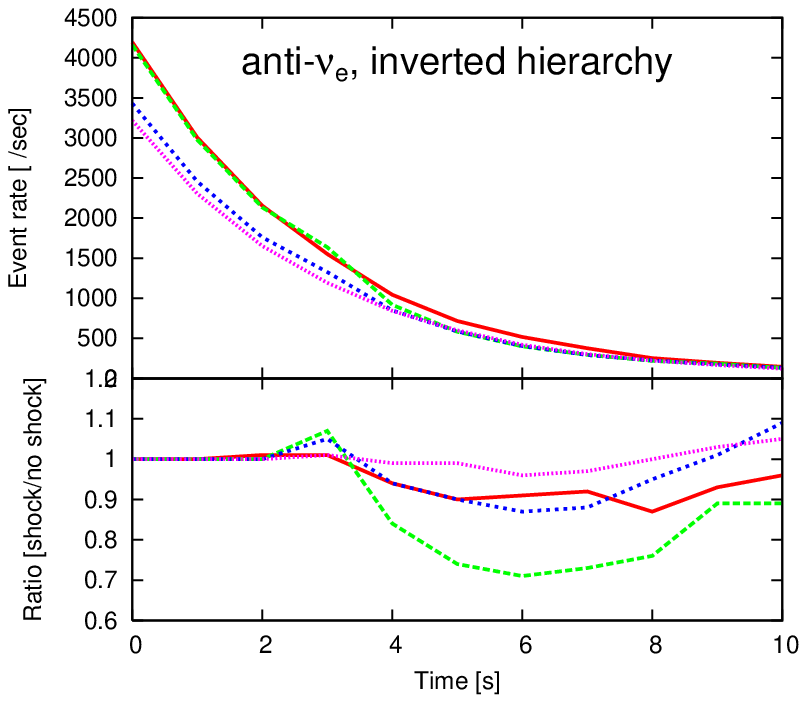}}
  \end{tabular}
 \end{center}
 \caption[The event eb of normal hierarchy.]
{\label{fig:eventrate-eb}
The same as Figure \ref{fig:eventrate-e}, but for $\bar{\nu}_e$. 
}
\end{figure}

%
Figure \ref{fig:eventrate-eb} is the same as Figure
\ref{fig:eventrate-e}, 
but for $\bar{\nu}_e$.
In the case of the normal hierarchy (left panel of Figure
\ref{fig:eventrate-eb}),
the event rate and the shock wave effect of $\bar{\nu}_e$ are
insensitive to  $\theta_{13}$ throughout the time profile, and
event rate decreases from about 3,200 at $t=0$ s to about 100 at
$t=10$ s.
This is attributed to the fact that
the adiabaticity of $\bar{\nu}_e$ is not influenced by 
$\theta_{13}$ or the shock wave as discussed in the previous section.

In the case of the inverted hierarchy (right panel of Figure
\ref{fig:eventrate-eb}), 
the behavior of event rate of $\bar{\nu}_e$ is different from $\nu_e$ in
normal hierarchy. 
At $t=0$ s, the event rate is about 4,200 for 
$\sin ^2 2 \theta_{13}=10^{-2}$ and $10^{-3}$, and it is about
3,200$-$3,500 for $\sin ^2 2\theta_{13}=10^{-5}-10^{-4}$.   
The event rate for large $\theta_{13}$ 
(i.e. $\sin ^2 2 \theta_{13}=10^{-2}$ and $10^{-3}$)
 is larger than that for
small $\theta_{13}$ (i.e. $\sin ^2 2 \theta_{13}=10^{-4}$ and
$10^{-5}$). 
When H-resonance is adiabatic, almost all $\bar{\nu}_e$ 
are completely converted from $\bar{\nu}_x$.
On the other hand, when H-resonance is non-adiabatic,
some $\bar{\nu}_e$ remain as $\bar{\nu}_e$ even after the neutrinos 
get out of 
the star.
As the result, the fraction of $\bar{\nu}_e$ converted from $\bar{\nu}_x$ is
small. 
The average energies of the three flavor neutrinos just emitted from the
proto-neutron star 
satisfy the following relation;
$\bar{E}_{\nu_e} < \bar{E}_{\bar{\nu}_e} <  \bar{E}_{\nu_x} $.
Therefore, the average energy of $\bar{\nu}_e$ in adiabatic case
becomes higher than that in non-adiabatic case.
As the main reaction cross section for 
$\bar{\nu}_e$ is
proportional to $E_{\nu}^2$ as discussed in section \ref{sec:level2c},
the detection probability of  high-energy neutrinos is larger than
that  of low-energy neutrinos.
Therefore, the event rate of $\bar{\nu}_e$ for large $\theta_{13}$
(adiabatic case) becomes much larger than that for small $\theta_{13}$
(non-adiabatic case).
This is the reason for large split between the expected event rates for
large $\theta_{13}$ (i.e. $\sin ^2 2 \theta_{13}=10^{-2}$ and $10^{-3}$)
and small $\theta_{13}$ (i.e. $\sin ^2 2\theta_{13}=10^{-4}$ and
$10^{-5}$).
We find
difference of event rates  by $\sim 1,000$ at $t = 0$ between the two
cases. 

Furthermore, the ratio of the event rates for $\bar{\nu}_e$ which
exhibits 
the shock wave propagation effect  
 changes dramatically 
when $\sin ^2 2 \theta_{13}$ is large (right lower panel of Figure \ref{fig:eventrate-eb}).
The ratio of $\bar{\nu}_e$
increases at around 2 s,
and decreases after 3 s, which is similar to $\nu_e$ event rates in
normal hierarchy (left lower panel of Figure \ref{fig:eventrate-e}).
The mechanism is quite similar to each other so that the shock wave
propagation through the H-resonance starts from low energy neutrinos to
high energy neutrinos, which eventually decreases the event rates.
However, the decrease in the ratio of the $\bar{\nu}_e$ event rates is
remarkable by at most 30 \%
for $\sin ^2 2 \theta_{13}=10^{-3}$.
On the other hand, the ratio changes weakly for the other $\theta_{13}$
values 
because H-resonance is non-adiabatic regardless of the
shock wave, although its degree is still larger than the $\nu_e$ event
rate (see left lower panel of Figure \ref{fig:eventrate-e}).

The event rate of $\bar{\nu}_e$ is extremely larger than that of $\nu_e$ 
(for a given neutrino oscillation parameter set).
This is mainly because 
the total cross section of $\bar{\nu}_e$-induced reactions is about
$10^2$ 
times larger than that of $\nu_e$.
Moreover, the total cross section of $\bar{\nu}_e$-induced  reactions is 
almost proportional to the 
square of neutrino energy, but the total cross section of $\nu_e$-induced
reactions is linearly proportional to the neutrino energy.
Owing to this  different energy dependence of the cross sections, 
one can expect that 
the effect of the shock wave should be identified more clearly in the 
$\bar{\nu}_e$ events.

In practice, since the shock effect is folded in the observed event
rates of $\nu_e$ and $\bar{\nu}_e$, we should carry out careful
theoretical analysis.
The observed event rates are statistically uncertain of the order of
their square root.
As for $\nu_e$ in normal hierarchy, the statistical error is the same
order as the change in the ratio $\sim 15 \%$ of the event rates with
and without the shock effect because absolute number of expected event
rates are small
$\frac{dN}{dt} \lesssim 70$ (Figure \ref{fig:eventrate-e}):
Actually, $\frac{dN}{dt} \sim 10$ at $t \sim 6$ s when the shock effect
becomes maximum strength, and thereby 
$\sqrt{\frac{dN}{dt}}/\frac{dN}{dt} \sim 0.3$.
On the other hand, for $\bar{\nu}_e$ in inverted hierarchy, the
statistical error is expected to be small enough because of large
absolute number of expected event rates
$\frac{dN}{dt}\lesssim 4000$ (Figure \ref{fig:eventrate-eb}): In this
case, since $\frac{dN}{dt}\sim 500$ at $t \sim 6$ s, 
$\sqrt{\frac{dN}{dt}}/ \frac{dN}{dt}\sim 0.05$ which is much smaller
than the maximal change in the ratio $\sim 30 \%$ of the event rates.

\subsection{Ratio of high- to low-energy neutrino events}
\label{high-low}

There is a potential difficulty in identifying 
the influence of the
shock wave upon the observed neutrino events 
because time evolution of the original 
neutrino spectra, which is unaffected by the shock wave 
as well as the MSW effect, is unknown. 
The original neutrino spectra must be assumed  theoretically a
priori. 
We therefore look for another useful observable that should 
show a clear
signature of the shock wave effect even when we do not know the time
evolution of the original neutrino spectra.

\subsubsection{Time-integrated ratio of neutrino events}

We consider now 
the ratio of the high-energy component to the
low-energy component of time-integrated neutrino events.
This ratio $R_{\mathrm{x}}$
is defined by 
\cite{Takahashi},
\begin{eqnarray}
R_{\mathrm{x}} 
\equiv \frac{\displaystyle
\int^{10\mathrm{s}}_{0\mathrm{s}} \int^{60 \mathrm{MeV}}_{20 \mathrm{MeV}}
\frac{d^2 N}{dE_e dt}dE_e dt}   
{\displaystyle \int^{10\mathrm{s}}_{0\mathrm{s}} 
\int^{20 \mathrm{MeV}}_{E_{\mathrm{th}}}
\frac{d^2 N}{dE_e dt}dE_e dt},
\end{eqnarray}
where x refers to SK or SNO and $E_{\mathrm{th}} = 7$MeV (SK) and 5 MeV
 (SNO) as explained below Eq. (\ref{eq:integral-event}). 
We set here the boundary between high- and low-energy components 
to be 20 MeV because 
the original energy spectra satisfy 
the conditions 
$\phi _{\nu_e}^{\mathrm{org}}=\phi_{\nu_x}^{\mathrm{org}}$ and 
$\phi _{\bar{\nu}_e}^{\mathrm{org}}=\phi_{\nu_x}^{\mathrm{org}}$
at
$E_{\nu}=$ 18.7 MeV and 23.3 MeV, respectively.

\begin{figure}
 \begin{center}
  \begin{tabular}{cc}
  \includegraphics[height=16pc, width=16pc]
  {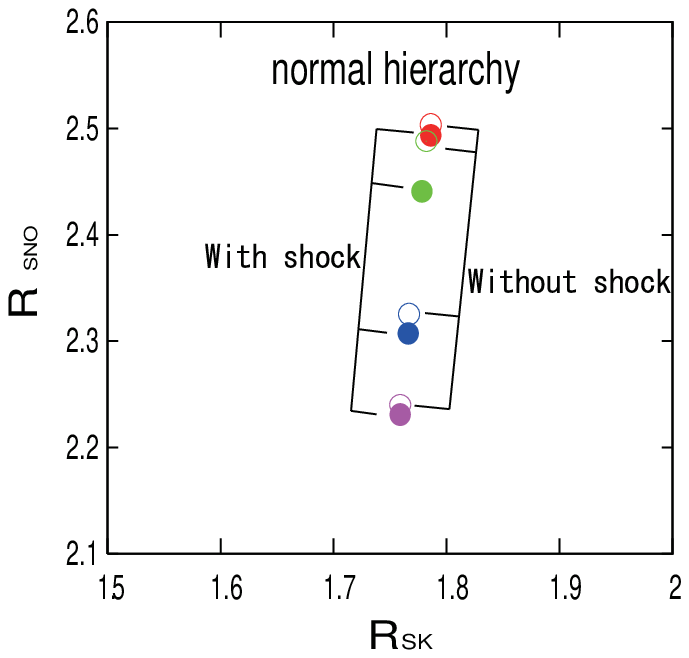}&
  \includegraphics[height=16pc]
  {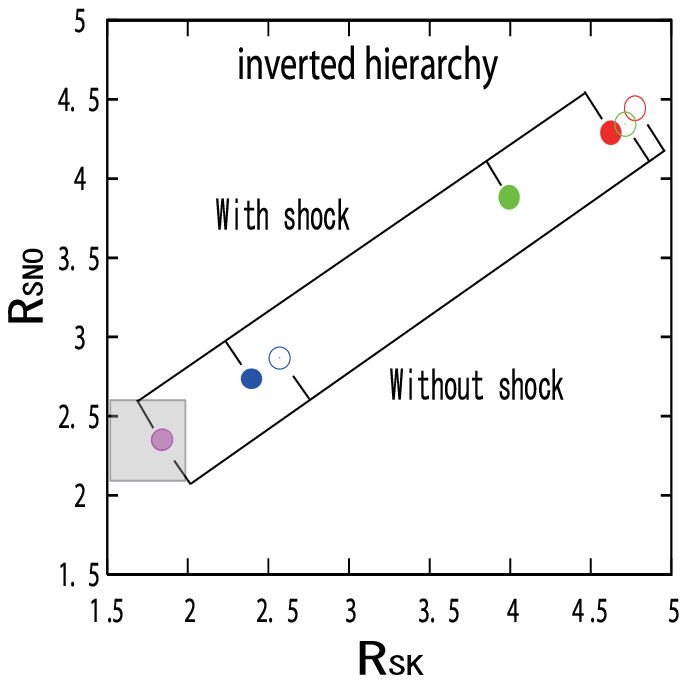}
  \end{tabular}
 \end{center}
     \caption[Comparison of the event in SK with SNO ]
{\label{fig:SK-SNO}Calculated correlation between $R_{\mathrm{SK}}$ and
 $R_{\mathrm{SNO}}$, which are 
the ratio of the time-integrated events of
 high-energy neutrinos (20 MeV $< E_{\nu} <$ 60 MeV) to those of
 low-energy neutrinos (5 MeV $< E_{\nu } <$ 20 MeV) 
for the SK and SNO 
detectors, respectively,
for various
 mixing angles of $\sin ^2 2 \theta_{13}=10^{-2}$(red), $10^{-3}$(green),
 $10^{-4}$(blue) and $10^{-5}$(purple).
Left panel is for normal hierarchy, and right panel for inverted
 hierarchy.  
Closed and open circles are the calculated results with and without
 shock  wave, respectively, as denoted.
Note that the scale of each panel is different, and a closed square in
 the right panel corresponds to a range in the left panel.
} 
\end{figure}

Figure \ref{fig:SK-SNO} shows $R_{\mathrm{SK}}$ vs $R_{\mathrm{SNO}}$
for various mixing angles 
$\sin ^2 2 \theta_{13}=10^{-2}$(red), $10^{-3}$(green),
 $10^{-4}$(blue) and $10^{-5}$(purple).
Left panel of Figure \ref{fig:SK-SNO} 
is in normal hierarchy, and right panel is in inverted
 hierarchy.  
Closed and open circles are the calculated results with and without
 shock  wave effect, respectively.
Note that the scale of each panel is different.

In the inverted hierarchy,
we see large variations of the ratios; $R_{\mathrm{SK}}$ $\sim$ 1.8 - 4.6, and
$R_{\mathrm{SNO}}$ $\sim $  2.3 - 4.3. 
In addition, 
$R_{\mathrm{SK}}$ and $R_{\mathrm{SNO}}$ show a clear correlation 
because the most dominated reactions for SK and SNO detectors are the
$\bar{\nu}_e$-induced charged current reactions; 
$\bar{\nu}_e + p \rightarrow e^+ + n$ for SK and 
$\bar{\nu}_e + d \rightarrow n + n + e^+$ for SNO.
Both ratios $R_{\mathrm{SK}}$ and $R_{\mathrm{SNO}}$ increases
with the mixing angle 
$\theta_{13}$.
For a given $\theta_{13}$, the ratios in the case with shock are smaller
than those without the shock.
On the other hand, in the normal hierarchy,  
the ratio shows a small dependence on $\sin ^2 2 \theta_{13}$ for
$R_{\mathrm{SK}}$ and $R_{\mathrm{SNO}}$.
In practice, we see only small variation of the ratios;
$R_{\mathrm{SK}}$ $\sim $ 1.76 - 1.8, and $R_{\mathrm{SNO}}$ $\sim $ 2.2
- 2.5. 
%
%
%
%
These characteristics in the parameter dependence of 
$R_{\mathrm{SK}}$ and $R_{\mathrm{SNO}}$ 
are consistent with the result of ref. \cite{Takahashi}.

The parameter dependence which we confirmed in $R_{\mathrm{SK}}$ and 
$R_{\mathrm{SNO}}$ as discussed above could be an important
observational signature to discriminate neutrino mass hierarchy, normal
or inverted, as well as the mixing angle, $\theta_{13}$.
In normal hierarchy, the ratios $R_{\mathrm{SK}}$ and $R_{\mathrm{SNO}}$
are small 
regardless of the mixing angle $\theta_{13}$ and the shock effect.
Furthermore, $R_{\mathrm{SNO}}$ changes a little according to the value of
$\theta_{13}$ and $R_{\mathrm{SK}}$ hardly changes.
We defined 
$R_{\mathrm{x}}$ as the total events of all flavors.
Most of the events detected in 
SK is $\bar{\nu}_e$.
As a result, the variation of $\nu_e$ events is hindered 
by larger  $\bar{\nu}_e$ events unchanged by the $\theta_{13}$
variation.  
On the other hand, the events of SNO contain the events of $\nu_e$ as
much as that of $\bar{\nu}_e$, because the cross section of
$\nu_e$-induced reaction is of the same order as that of
$\bar{\nu}_e$-induced reaction.
Therefore, we can see the variation of $\nu_e$.
However, the variation of $\nu_e$ is small because 
the number of events is not so different between $\nu_e$ and
$\bar{\nu}_e$.
In inverted hierarchy, the ratios $R_{\mathrm{SK}}$ and
$R_{\mathrm{SNO}}$  correlate with 
the mixing angle $\theta_{13}$ and the shock effect. 
Larger $\theta_{13}$ value indicates larger $R_{\mathrm{SK}}$ and
$R_{\mathrm{SNO}}$ 
ratios. 
However, the shock effect reduces $R_{\mathrm{SK}}$ and
$R_{\mathrm{SNO}}$ even  in large
$\theta_{13}$ value.
Therefore, we do not distinguish the value of $\theta_{13}$ and the
shock effect in $R_{\mathrm{SK}}$ and $R_{\mathrm{SNO}}$.


\subsubsection{Time-dependent ratio of neutrino events}
In the previous subsection, we discussed the ratio of high-energy to
low-energy events, $R_{\mathrm{x}}$, which is the ratio of the
neutrino events integrated over the time.
We also evaluated in Sec. \ref{sec:eventrate}
the time-dependent event rates integrated over the neutrino energy.
Figures \ref{fig:eventrate-e}
and \ref{fig:eventrate-eb} indicate that some effects of
shock wave propagation could be observed at later times when the shock
wave reaches H-resonance.
However, 
we apparently lose some important information on the effects of shock
wave in either case.
In this subsection, we explore for the signature of the shock wave
effects by taking account of double differential as to both time and
energy.

\begin{figure}
 \begin{center}
   \resizebox{165mm}{!}{\includegraphics{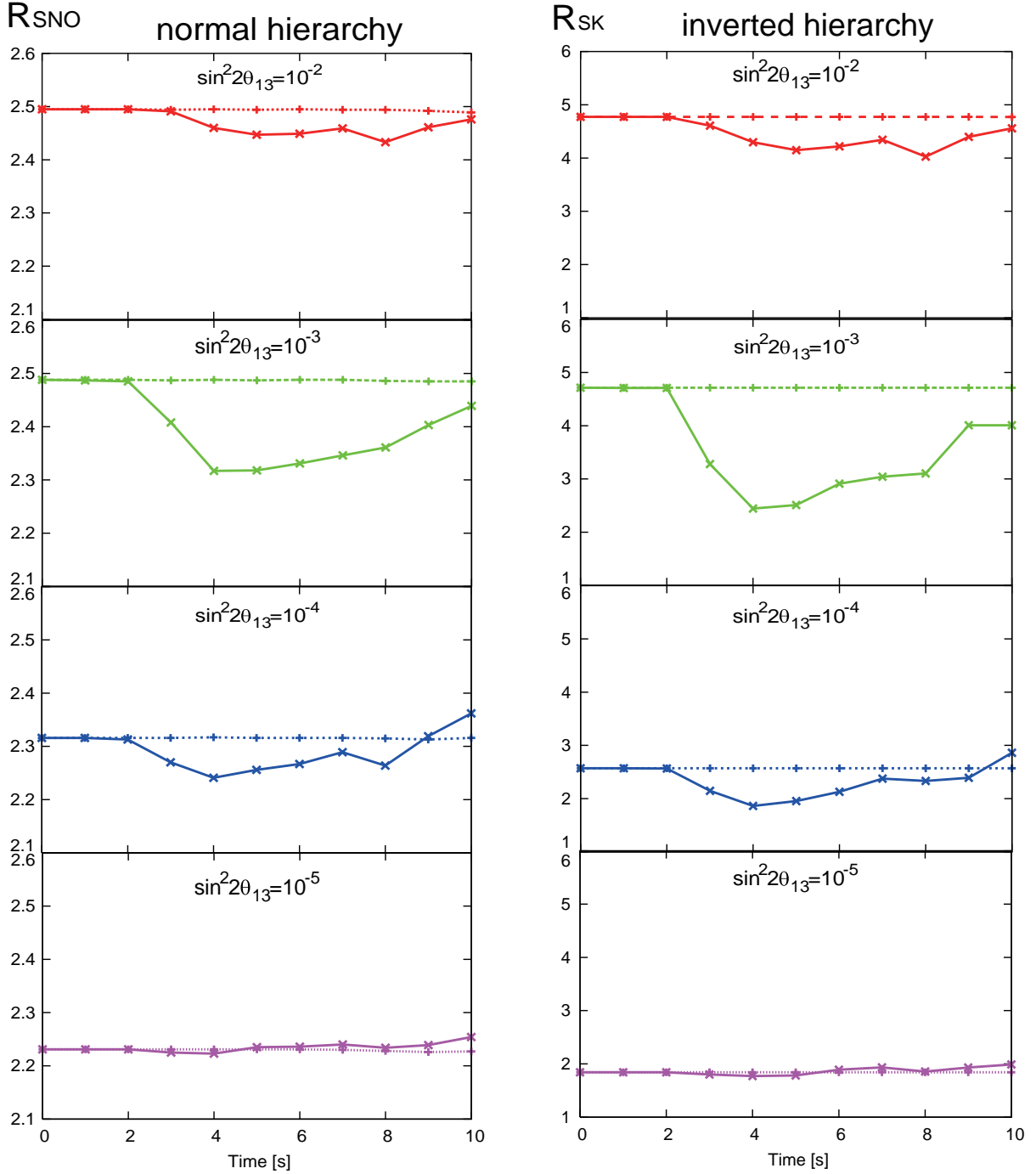}}
 \end{center}
 \caption{\label{fig:timeratio}
Time-dependent ratios of the events of high-energy neutrinos 
(20 MeV $<  E_{\nu} <$ 60 MeV) to those of low-energy neutrinos
(5 MeV $< E_{\nu} <$ 20 MeV) for the SK and SNO detectors for various
 mixing angles
of  $\sin ^2 2\theta _{13}$ 
$=10^{-2}$, $10^{-3}$, $10^{-4}$ and
 $10^{-5}$ from top to bottom panel.
Left panels are the ratio $R_{\mathrm{SNO}}(t)$ for normal hierarchy,
 and right panels
 are the ratio $R_{\mathrm{SK}}(t)$ in inverted hierarchy.
Solid and dashed lines are the calculated results with and without shock
 wave, respectively.
Note that the scale of each panel is different.
}
\end{figure}

We here define time-dependent ratio of the events of high-energy to
low-energy neutrinos, 
\begin{eqnarray}
R_{\mathrm{x}}(t)
\equiv \frac{\displaystyle \int^{60 \mathrm{MeV}}_{20 \mathrm{MeV}}
\frac{d^2 N}{dE_e dt}dE_e }
{\displaystyle \int^{20 \mathrm{MeV}}_{E_{\mathrm{th}}}
\frac{d^2 N}{dE_e dt}dE_e },
\end{eqnarray}
where x refers to SK or SNO.
Figure \ref{fig:timeratio} shows this ratio
$R_{\mathrm{x}}(t)$ 
as a function of time.  
The values of  $\sin ^2 2\theta _{13}$ 
are $10^{-2}$, $10^{-3}$, $10^{-4}$ and
 $10^{-5}$ from top to bottom panels.
Left panels are for
normal hierarchy
and right panels 
 are for 
inverted hierarchy.
Solid and dashed lines are the calculated results with and without shock
 wave, respectively.
Note that the scale of each panel is different.

First, we show $R_{\mathrm{SNO}}(t)$ in normal hierarchy
in left panels of Figure \ref{fig:timeratio}.
The effect of shock wave is seen, 
but it is a
very small effect of at most about 10 \%.
As for observed $\nu_e$ spectrum in normal hierarchy, 
70 \% is the original $\nu_x$ and 30
\% is the original $\nu_e$ in the non-adiabatic case, though almost 100
\% is the original $\nu_x$ in the adiabatic case.
On the other hand, 70 \% of the observed $\nu_e$ is the original
$\bar{\nu}_e$ and 30 \% is the original $\nu_x$ in inverted hierarchy. 
Moreover,  the shock effect of $\nu_e$ in normal hierarchy is washed out by
neutral current and charged current of $\bar{\nu}_e$ because the target
of neutrino detection is deuteron.
%
Therefore, the variation of the event rate ratio is small in the normal
hierarchy. 
There is, however, possibility of finding shock wave effect if much
larger events are 
detected by the
next
generation detector. 

Second, we show $R_{\mathrm{SK}}(t)$ in inverted hierarchy
in right panels of Figure \ref{fig:timeratio}.
The event rate ratio $R_{\mathrm{SK}}(t)$ is constant  in
early times and strongly depends on $\sin ^2 2 \theta_{13}$.
It 
is 4.7 and 1.8 in the case of 
$\sin ^2 2 \theta_{13}=10^{-3}$ and $10^{-5}$, respectively, at $t=0$
s. 
The value of $R_{\mathrm{SK}}(t)$ 
is large when 
$\sin ^2 2 \theta_{13}$ is large, because
the H-resonance is adiabatic and the average energy of $\bar{\nu}_e$ 
is high.  
The small $R_{\mathrm{SK}}(t)$
value is due to non-adiabatic state of H-resonance. 

When the shock wave effects are included and  $\sin^2 2 \theta _{13}$ is
$10^{-3}$ (large),  
$R_{\mathrm{SK}}(t)$ greatly changes with time.
Although 
$R_{\mathrm{SK}}(t)$ is 4.7 until 2 s, it decreases to 
$\sim 2$, which is closer to the ratio in the case of non-adiabatic
state in 4 - 8 s.
This decrease is due to the change of 
the adiabaticity of H-resonance according to the shock propagation.
The H-resonance is adiabatic before
the shock arrival to the resonance.
In contrast, the H-resonance is non-adiabatic while the
shock wave propagates in the resonance, and 
$R_{\mathrm{SK}}(t)$
approaches $R_{\mathrm{SK}}(t) \sim 1.8$ which is the ratio
in small $\sin^2 2\theta_{13}$ case (see the bottom panel).

When $\sin^2 2\theta _{13}$ is as small as $10^{-5}$, 
we see only small differences of
$R_{\mathrm{SK}}(t)$ between the cases 
with and without the shock wave 
and hardly see the time dependence.
The H-resonance is
non-adiabatic whenever the shock wave is on the resonance
or not in this case of small $\sin ^2 2\theta_{13}$.

To conclude, we can see  a clear signature of 
the shock wave in the time-dependent high-energy to low-energy ratio
of $R_{\mathrm{x}}(t)$ although 
the influence of the shock wave is less clearly seen in the ratio of 
time integrated event rates, $R_{\mathrm{x}}$. 
Therefore, the observations of the time evolution of 
$R_{\mathrm{x}}(t)$ is 
quite an important observable to constrain the neutrino oscillation
parameters and  to find
the effects of shock propagation in supernovae.

\subsection{\label{sec:level4}Further discussion}

Fogli et al.\cite{Fogli2}
 found qualitatively similar result to ours in the
calculated event rates with and without forward shock in the case of
inverted hierarchy for $\sin^2 2 \theta_{13} = 10^{-2}$.  Their results are
however different from ours quantitatively for several reasons.  For
one, their event rates at the neutrino energy 45 $\pm$ 5 MeV are
calculated by assuming the next generation detector of 0.4 Mton pure
water that is 12.5  bigger than 32 kton for SK.  We can scale our
calculated results by 12.5 times in order to remove this apparent
difference.  The second reason is that they adopted completely different
supernova model from ours.  Although it is hard to reconcile the
difference between the two models, we tried to compare at least the
effect of shock wave propagation in the following manner.

We should, first, refer to our result of the event numbers at the same
neutrino energy 45 $\pm$ 5 MeV at 5 s after the core bounce for
$\sin^2 2 \theta_{13} = 10^{-3}$.  This aims to compare the two results of the
event rates so that the shock wave effect appears most remarkably in
both calculations in the case of inverted hierarchy.  Second, we should
remove the bias based on different supernova models especially arising
from different density profiles from each other.  For this purpose we
normalize our event number without shock to Fogli's event number without
shock.  We thus obtain finally the following event numbers; 135 with
shock and 300 without shock in inverted hierarchy.  These scaled and
normalized numbers should be compared with Fogli's results; 130 with
shock and 300 without shock.  Note again that the latter number in
either calculation is the same by definition of normalization.  Our
numerical calculations show 55.0 \% decrease in the event number as a
resultant net effect of shock wave propagation, which is in reasonable
agreement with 56.7 \% decrease as shown by their 
calculations.  
We thus
conclude that their 
theoretical calculations and ours of the neutrino
event rates agree with each other despite several essential differences:
 We solved in Eq. (1) both flavor conversion of neutrinos and supernova
density profile numerically in our method described in Sec. 1, while
Fogli et al. \cite{Fogli2}
 applied an artificial model of density profile of
shock wave propagation without solving dynamical supernova explosion.

For further comparison, 
let us analytically estimate the decrease in the neutrino events in the
non-adiabatic case (with shock) compared with the adiabatic case
(without shock).
In the non-adiabatic case, we define the survival probability of
$\bar{\nu}_e$  as $\bar{P}=0.7$ in our model, and we use the result of
$\bar{\nu}_e$ in the case of normal hierarchy in Fogli's  model.
We set $\bar{P}=0$ in the adiabatic case.
Thus estimated analytical results show 62.3 \% and 63.0 \% decrease in
the neutrino events from adiabatic to non-adiabatic cases in our model
and their model, respectively.
These values are close enough to each other, and $\sim 60 \%$ are not
very different from $\sim 55.0\%$ which was inferred  from numerical
calculations as discussed in the previous paragraph.
This fact justifies that our treatment of analytical estimates are
reasonable. 
Therefore, regardless of all possible differences between the two
models,  
the supernova neutrinos provide a powerful tool to indicate 
the shock wave propagation  inside  supernova 
if $\sin ^2 2 \theta_{13}$ is large in the inverted hierarchy.

We assumed that CP violating phase is zero
and $\sin ^2 2 \theta_{23}=1$ throughout this study.
However, CP-phase is unknown 
and there is an uncertainty in the 
mixing angle $\theta_{23}$.
Here we discuss the dependence of the shock effect on neutrino flavor transitions
on CP-phase and $\theta_{23}$.
Transition probabilities of neutrinos in arbitrary density profile
have been studied theoretically
in \cite{yokomakura2002},
where it was found
that the transition probabilities of 
$\nu_e \rightarrow \nu_e$
and $\bar{\nu}_e \rightarrow \bar{\nu}_e$ do not depend on CP-phase and
mixing angle $\theta_{23}$ at all.
Furthermore, 
one can show 
from their formula in \cite{yokomakura2002} that 
the sum of the transition probabilities of 
$\nu_{\mu} \rightarrow \nu_{e}$ and $\nu_{\tau} \rightarrow \nu_{e}$ 
($\bar{\nu}_{\mu} \rightarrow \bar{\nu}_{e}$ and 
$\bar{\nu}_{\tau} \rightarrow \bar{\nu}_{e}$)
are totally free from
CP-phase and $\theta_{23}$.
Since the spectra of $\nu_{\mu}$ and $\nu_{\tau}$ 
($\bar{\nu}_{\mu}$ and $\bar{\nu}_{\tau}$) are the same at the neutrino 
sphere, the transition probability from $\nu_{\mu}$ or
$\nu_{\tau}$ ($\bar{\nu}_{\mu}$ or $\bar{\nu}_{\tau}$) to
$\nu_e$ ($\bar{\nu}_e$) can be written as a half of 
the sum of the transition probabilities of 
$\nu_{\mu} \rightarrow \nu_e$
and $\nu_{\tau} \rightarrow \nu_e$ 
($\bar{\nu}_{\mu} \rightarrow \bar{\nu}_{e}$ and 
$\bar{\nu}_{\tau} \rightarrow \bar{\nu}_{e}$).
The transition probability from $\nu_{\mu}$ or
$\nu_{\tau}$ ($\bar{\nu}_{\mu}$ or $\bar{\nu}_{\tau}$) to 
$\nu_e$ ($\bar{\nu}_e$) does not depend on CP-phase and $\theta_{23}$.
We thus conclude that there is no effect of CP-phase and mixing angle $\theta_{23}$ 
on the supernova neutrino spectra.

If the influence of the shock wave is seen very early ($t \leq 1 s$)
in the observation
of the supernova neutrinos, it might mean that the shock wave reaches
H-resonance ($\sim 1000 \; \mathrm{g/cm}^3$) very early after core bounce.
Then, we would expect that this kind of supernova explosion exhibits a
strong shock wave effect in the direction of observer.
Moreover, assume that we could know viewing angle 
from the axis of a supernova in
optical and radio observations.
If the viewing angle 
is small
and
the influence of the shock wave is seen very early, this supernova has
strong shock like a narrow beaming jet along the axis of supernova
explosion. 
Therefore, we would expect that this supernova explodes in the mechanism
associated with the jet formation 
\cite{Obergaulinger,Uzdensky,Moissenko,Burrows07,Sawai,Mikami,Takiwaki09}. 
If the viewing angle is large
and the influence of the
shock is seen very early, we would expect that this supernova has a
rather wide
jet and may be close
to the spherical symmetric explosion.

On the other hand, if the influence of the shock wave is seen in the
supernova neutrinos at relatively later time, 
and the viewing angle is small,
we would expect that this
explosion is also close to the spherical symmetry without jet. 
Therefore, we would expect that this supernova explodes in the mechanism 
without a jet. 
Details of the influence of neutrino oscillation on jet explosion are
discussed in \cite{Kawagoe09}.

Recently, oxygen emission-line profiles from supernovae are observed,
which could be a signature of an aspheric explosion \cite{Maeda08}.
We would study the asphericity of the supernova explosions more
in detail 
from such an optical observation and simultaneous detection
of the supernova neutrinos.  

The flavor-exchange effects by the neutrino self-interactions
might also be important because of 
their huge flux immediately after the emission out of the
proto-neutron star (e.g., \cite{Fogli07, Duan09}).  
This could change initial neutrino spectrum from what we assumed here. 

\section{\label{sec:level5}Conclusion}

We investigated the neutrino signal to study the effect of the
shock wave propagation 
as well as the dependence on unknown neutrino oscillation parameters of
mass hierarchy and $\theta_{13}$ by carrying out numerical calculations
of the neutrino event number
using the simulation result of a supernova explosion.
We followed adiabatic collapse of iron core, core bounce, and the shock
wave propagation in the stellar envelope for a  long time
( $\gtrsim 10$ s) using general 
relativistic hydrodynamical code and
realistic density profile. 
Then, we could calculate detailed 
time
evolution of the event number rate and spectra of the supernova neutrinos.

The expected event rate of
$\bar{\nu}_e$ in the case of the inverted mass hierarchy 
and that of $\nu_e$ in the normal mass hierarchy 
depend on the mixing angle $\theta_{13}$.
When $\sin^22\theta_{13}$ is larger than $\sim 10^{-4}$, 
the influence of the shock wave
appears in the observation after 2 s in our model.
It is the time when the shock wave reaches the H-resonance.
Therefore, the shock effect and the constraint on $\theta_{13}$ can be
inferred  even
by the event rate integrated by the whole energy range. 
However, it is  difficult to distinguish the influence of the
shock wave and the neutrino oscillation parameters only from the time
evolution of 
the event rate.

The time-integrated  ratio of the events of high-energy to
low-energy neutrinos is another indicator to examine the shock effect
and constrain mass hierarchy and $\theta_{13}$.
The parameter dependence of the time integrated ratio $R_{\mathrm{x}}$ 
(x = SK or
SNO) 
could be an
important observational signature to discriminate neutrino mass
hierarchy, normal 
or inverted, as well as the mixing angle, $\theta_{13}$.
Both of the ratios $R_{\mathrm{SK}}$ and $R_{\mathrm{SNO}}$
 correlate with the value of
$\theta_{13}$ in inverted mass hierarchy.
Therefore, the neutrino oscillation parameters $\theta_{13}$ and
the mass hierarchy, might be able to be constrained by this ratio.
The shock effect reduces the ratios $R_{\mathrm{SK}}$ and
$R_{\mathrm{SNO}}$.
This effect is not distinguished with smallness of $\theta_{13}$. 

The time-dependent ratio of high- to low-energy neutrino events is more
preferable indicator to find out the shock effect clearly.
The ratio decreases after 2 s for $R_{\mathrm{SNO}}(t)$
 in normal hierarchy and
for $R_{\mathrm{SK}}(t)$
in inverted hierarchy.
The decrease is the most remarkable in the case of 
$\sin ^2 2\theta_{13}=10^{-3}$ and more clearly in inverted hierarchy.
The dramatic decrease in the ratio 
could be a clear signal for 
the shock wave effect and would constrain the minimum value of
 $\sin^2 2\theta_{13}$. 
Therefore, observations of  
the time-dependent ratio of the high- to low-energy neutrino event
are an
important observable to find the effects of shock propagation in
supernovae.  

\section*{Acknowledgments}
We would like to  thank K. Kotake for fruitful discussions. 
SK thanks to T. Takiwaki and S. Harikae for help comments 
on the numerical scheme.
This research work was supported in part by Research
Fellowships of the Japan Society for the Promotion of Science (JSPS).


\end{document}